\documentclass[fleqn,usenatbib]{mnras}

\usepackage{newtxtext,newtxmath}

\usepackage[T1]{fontenc}

\DeclareRobustCommand{\VAN}[3]{#2}
\let\VANthebibliography\thebibliography
\def\thebibliography{\DeclareRobustCommand{\VAN}[3]{##3}\VANthebibliography}

\usepackage{graphicx}	
\usepackage{amsmath}	
\usepackage{bm}
\usepackage{comment}
\usepackage{hyperref}

\newcommand\LCDM{$\Lambda$CDM}

\title[WL convergence in screened MG theories]{Ray-traced weak lensing convergence in screened modified gravity theories}

\author[M. Pantiri et al.]{
Mattia Pantiri,$^{1}$\thanks{E-mail: pantiri@lorentz.leidenuniv.nl}
Matthieu Schaller,$^{1,2}$
Alessandra Silvestri,$^{1}$ Jeger C. Broxterman$^{1,2}$ and Joop Schaye$^{2}$
\\
$^{1}$Lorentz Institute for Theoretical Physics, Leiden University, PO Box 9506, NL-2300 RA Leiden, the Netherlands\\
$^{2}$ Leiden Observatory, Leiden University, PO Box 9513, NL-2300 RA Leiden, the Netherlands\\
}

\date{Accepted XXX. Received YYY; in original form ZZZ}

\pubyear{\the\year{}}

\begin{document}
\label{firstpage}
\pagerange{\pageref{firstpage}--\pageref{lastpage}}
\maketitle

\begin{abstract}
Weak gravitational lensing is one of the primary cosmological probes, providing powerful constraints on the cosmological model. As Stage IV surveys are expected to deliver data of unprecedented precision, accurate modeling of weak gravitational lensing observables across both linear and non-linear scales becomes increasingly important. 
In this work, we investigate weak lensing in modified gravity (MG) models, extensions of the standard \LCDM{} cosmology in which gravity deviates from general relativity, generally introducing modifications to the lensing equation. We parametrize these modifications through the common phenomenological function $\Sigma_\mathrm{mg}$   and apply ray-tracing to the density maps of N-body and  hydrodynamical simulations.  We model the time dependence of $\Sigma_\mathrm{mg}$ analytically, while we introduce a phenomenological scale dependence to represent the screening mechanisms by which MG models reduce to  general relativity in high-density environments.
Starting from the output of the FLAMINGO hydrodynamical simulations, we generate fully ray-traced convergence maps using our modified lensing model. We analyze how the parameters of our prescription affect the weak lensing convergence power spectrum and compare these effects to other known sources of variation, in particular cosmological parameters and baryonic feedback.
We find that the modifications to the lensing equation deriving from the MG model produce non-negligible signatures in the convergence power spectrum and that, within extensions of the \LCDM{} framework, these effects can be larger than those induced by baryonic physics. Our results indicate that modified lensing should become a standard ingredient of the analysis of modified gravity simulations.
\end{abstract}




\section{Introduction}
\label{sec:introduction}

The standard model of cosmology, known as the \LCDM{} model, has been remarkably successful in explaining a wide range of observations, from the anisotropies of the cosmic microwave background (CMB; e.g. \citealt{Planck_2020}) to the large-scale distribution of galaxies (e.g. \citealt{Anderson_2014}). Despite its great success, the advent of the precision cosmology era has revealed several tensions when combining independent datasets, such as the Hubble tension and the $S_8$ (or $\sigma_8$) tension (e.g. \citealt{Verde_2019, Di_Valentino_2021, Abdalla_2022}). More recently, \cite{DESI_2024VI} found a discrepancy of up to $3.5\sigma$ with \LCDM{} in the dark energy equation of state.

Moreover, many fundamental questions within the \LCDM{} model remain unanswered, such as the nature of dark matter and dark energy. These open issues have motivated the exploration of extensions to \LCDM{}, some of which question the validity of general relativity (GR) as the theory of gravity on cosmological scales. Although GR has been exquisitely tested in the Solar System (e.g. \citealt{Will_2014}), its validity on much larger scales has yet to be submitted to the same level of scrutiny. In many modified gravity (MG) frameworks, an additional scalar degree of freedom is introduced, which can provide a dynamical explanation for the late-time accelerated expansion of the Universe (e.g. \citealt{Clifton_2012}).

The scalar field contributes to the background dynamics, but in general can also cluster and alter the lensing, thus  imprinting a variety of features in the main observables of cosmological surveys (e.g. \citealt{Silvestri:2009hh}), such as the clustering of galaxies and weak gravitational lensing (WL) --- the latter being the focus of this work. For reviews of weak lensing, see, e.g., \cite{Kilbinger_2015} and \cite{Bartelmann_2001}. Weak gravitational lensing arises from the small deflections of light rays traveling through the inhomogeneous large-scale structure of the Universe, leading to a measurable distortion, or cosmic shear, in the observed shapes of galaxies. If the theory of gravity differs from GR, this affects both the clustering of matter and the lensing potential itself --- the key theoretical quantity from which WL observables, such as the convergence and shear, can be computed. 

At the linear level, significant effort has gone into studying the modifications to  both clustering and lensing resulting from MG. These have both been implemented, for instance, in modifications of Einstein-Boltzmann solvers, such as MGCAMB \citep{Zhao_2009, Wang_2023}, and MGclass \citep{Lewis_2000, Sakr_2022}. At the non-linear level, on the other hand, most of the focus has been on the modifications to the clustering, with MG models being studied via modified N-body simulations (e.g \citealt{Winther_2015}), emulators (e.g. \citealt{Saadeh_2024}), halo reaction methods \citep{Bose_2020}, and other hybrid methods (e.g. \texttt{Hi-COLA}, \citealt{Wright_2023}). The MG non-linear clustering, which can be derived from these approaches, can then typically be used in combination with the standard GR lensing prescription to get predictions for lensing, either via the matter power spectrum or via the ray-tracing procedure, depending on the availability of highly resolved density maps (see for instance \citealt{harnoisderaps_2022} and \citealt{hoyland2025fastgenerationweaklensing}). Very few attempts have been made at incorporating also a modified lensing prescription (e.g. \citealt{Barreira_2017}). This is partly justified by the fact that popular MG models such as f(R) and DGP modify clustering significantly, but lensing only very mildly, or not at all,  primarily via background effects (e.g. \citealt{Lue_2004, Pogosian_2016}). However, the gravitational landscape of MG models is vast, and the lensing equation generally displays non-negligible time- and scale-dependent modifications.  Precision cosmology requires an exquisite modeling of all aspects. In an effort complementary to the ongoing ones, in this paper we will focus on the modifications of lensing and adopt the common phenomenological  function $\Sigma_{\rm mg}$ to capture the modified relation between the lensing potential and the density of matter (e.g. \citealt{Zhang_2007, Amendola_2008, Daniel_2010}). We will make a first  step towards integrating them into the derivation of convergence maps. We will consider scenarios in which $\Sigma_\mathrm{mg}$ is purely a function of time and is unaffected by the clustering of the scalar field; in other words, it is not expected to be screened on small scales. However, we will also consider scenarios in which $\Sigma_\mathrm{mg}$ instead displays scale-dependent suppression, reducing to the GR value in highly dense regions. Lensing equations in the non-linear regimes of MG models have not been studied with the same level of scrutiny as the equations governing clustering, and little is known about how the non-linear regime will affect $\Sigma_\mathrm{mg}$.  These two scenarios can be seen as the two limiting cases, with most MG models expected to fall somewhere in between. We will address the implementation of modified clustering in simulations in future work. 

Ongoing and upcoming Stage IV cosmological surveys, such as  \textit{Euclid} \citep{Euclid_2025}, \textit{Rubin/LSST} \citep{lsstsciencecollaboration2009lsstsciencebookversion}, and \textit{Roman} \citep{spergel2015widefieldinfrarredsurveytelescopeastrophysics}, will measure WL observables with unprecedented precision, probing scales where both linear and non-linear effects are relevant. For this reason, it is essential to model small-scale phenomena accurately, including baryonic feedback and screening mechanisms. Screening is a key feature of viable MG models: it ensures that GR is recovered in high-density environments where the theory has been thoroughly tested. The specific realization of screening depends on the model, but it can occur mainly through four mechanisms: symmetron \citep{Hinterbichler_2010}, Vainshtein \citep{Nicolis_2009}, k-mouflage \citep{Babichev_2009}, and chameleon \citep{Khoury_2004}. Reviews on modified gravity and screening can be found in \cite{Joyce_2015} and \cite{Koyama_2016}. In this work, we implement a phenomenological description of screening rather than a detailed, model-specific prediction.

The analysis presented here focuses on the angular power spectrum of the WL convergence, obtained via a backward ray-tracing algorithm \citep{Broxterman_2024} without relying on the Born approximation\footnote{The Born approximation consists in evaluating deflections on the unperturbed light ray path, as presented in Section \ref{sec:theory}.}. The ray-tracing uses as input the mass maps from the FLAMINGO cosmological hydrodynamical simulation suite \citep{Schaye_2023, kugel2023flamingocalibratinglargecosmological}, which has been calibrated using machine-learning techniques to reproduce the observed galaxy stellar mass function at redshift $z=0$ and cluster gas fractions. The simulations assume GR as the theory of gravity, and we will modify the ray-tracing prescription of \cite{Broxterman_2024} to include MG effects. The large volume (greater or equal to 1 Gpc) of the FLAMINGO simulations, together with their modeling of the small-scale baryonic physics, makes them suitable for our purpose, despite the assumption of GR, which will be lifted in future works. Finally, each hydrodynamical simulation in FLAMINGO has a dark-matter-only (DMO) counterpart, allowing us to assess the relative importance of MG effects and baryonic physics on WL observables (see, e.g., \citealt{Semboloni_2011, Chisari_2018}). 

The paper is organized as follows: in Section \ref{sec:theory}, we review the main equations involved in WL theory, in particular in the framework of general MG models. We then introduce the FLAMINGO hydrodynamical simulations, the ray-tracing algorithm, and the implementation of screening in Section \ref{sec:methods}. In Section \ref{sec:results}, we present our results, and we finally draw our conclusions in Section \ref{sec:conclusions}.


\section{Weak Lensing in Modified Gravity}
\label{sec:theory}

Before delving into the ray-tracing method and its numerical implementation, we briefly discuss the main equations of weak lensing theory, highlighting the contributions coming from a non-standard theory of gravity. An extensive review on weak lensing theory within the \LCDM{} model can be found in \cite{Kilbinger_2015}. We will work in the perturbed FLRW metric (following the notation of \citealt{albuquerque2025euclid})
\begin{equation}
    \rm{d}s^2 = -(1+2\Psi)\rm{d}t^2 + a^2(t)(1 - 2\Phi)\delta_{ij}\rm{d}x^i \rm{d}x^j\,,
\end{equation}
where $\Phi$ and $\Psi$ are the Bardeen potentials \citep{bardeen_1980}. In the GR limit, they are equal to each other and equal to the standard Newtonian gravitational potential. Consider a light ray in this metric traveling from its source to the observer, located at the origin of the coordinate system. Its path will be affected by small deviations from a straight line due to the large-scale structure of the Universe. Each individual small angular deviation is called a \emph{deflection angle} $\bm{\alpha}$. We can then give an expression for the angular position $\bm{\beta}$ of the light ray at any comoving distance $\chi$ by integrating the deflection angle from the observer position up to $\chi$
\begin{equation}
    \bm{\beta}(\bm{\theta}, \chi) = \bm{\beta}(\bm{\theta}, 0) + \int_0^\chi \rm{d}\chi' \frac{\chi - \chi'}{\chi\chi'} \nabla_\perp\Phi_W(\bm{\beta}(\bm{\theta}, \chi'), \chi')\,,
    \label{eq:light_ray_position}
\end{equation}
where $\bm{\theta}$ is the observed angular position of the light ray and $\nabla_\perp\Phi_W$ is the gradient of the Weyl potential along the direction perpendicular to the light ray path. The \emph{Weyl potential} is defined as the average of the Bardeen potentials
\begin{equation}
    \Phi_W = \frac{\Phi + \Psi}{2}\,,
\end{equation}
and it is evaluated at the true angular position $\bm{\beta}$, meaning that we will not consider the Born approximation (for which we would evaluate the Weyl potential at the observed angular position $\bm{\theta}$). 
The main features relevant to WL can be summarized by the magnification matrix, defined as the derivative of $\bm{\beta}$ with respect to $\bm{\theta}$ and usually decomposed as
\begin{equation}
    A = \frac{\partial\bm{\beta}}{\partial\bm{\theta}} \approx 
    \begin{pmatrix}
        1 - \kappa - \gamma_1 & -\gamma_2 + \omega \\
        -\gamma_2 - \omega & 1 - \kappa + \gamma_1 \\ 
    \end{pmatrix}
    \,,
    \label{eq:mag_matrix}
\end{equation}
where $\kappa$ is the WL convergence - indicating the isotropic part of the distortion of the image - $\gamma = \gamma_1 + i\gamma_2$ is the shear - related to anisotropic distortions - and $\omega$ is the image rotation angle. We also define the \emph{lensing potential} as the projection of the Weyl potential
\begin{equation}
    \psi(\bm{\theta}, \chi) = \int_0^\chi \rm{d}\chi' \frac{\chi - \chi'}{\chi\chi'} \Phi_W(\bm{\beta}(\bm{\theta}, \chi'), \chi')\,.
\end{equation}
This is a key quantity in WL theory, because from its derivatives it is possible to recover both the convergence and the shear. 

In this work, we will focus on the convergence $\kappa$, which can be readily derived from the magnification matrix as
\begin{equation}
\label{eq:convergence_from_A}
    \kappa(\bm{\theta}, \chi) = 1 - \frac{A_{11} + A_{22}}{2}\,.
\end{equation}
The $\chi$-dependent convergence can then be integrated along the line of sight and convolved with a light source distribution $n(\chi)$,  for instance, matching the expected ones for specific Stage IV surveys. This leads to the convergence as an angular function on the sphere
\begin{equation}
\label{eq:convergence_integrated}
    \kappa(\bm{\theta}) = \int_0^\chi \rm{d}\chi' n(\chi') \kappa(\bm{\theta}, \chi')\,,
\end{equation}
which will be the main object that we study in this work. More specifically, we will consider its angular power spectrum which is the spherical harmonic transform of its correlation function.

In order to understand where the theory of gravity has an impact on the equations just presented, let us look at two fundamental equations: the Poisson equation and the lensing equation. In MG models, it is possible to capture the modifications to GR through two phenomenological functions $\mu_\mathrm{mg}$ and $\Sigma_\mathrm{mg}$, which enter on the right hand side of the aforementioned equations as follows \citep{Zhang_2007, Amendola_2008, Daniel_2010}
\begin{align}
    \nabla^2\Psi & = 4\pi G_\mathrm{N} \mu_\mathrm{mg}(a, k) a^2\rho_\mathrm{m}\delta_\mathrm{m} \,, \label{eq:clustering}
    \\
    \nabla^2\Phi_W & = 4\pi G_\mathrm{N} \Sigma_\mathrm{mg}(a,k)a^2\rho_\mathrm{m}\delta_\mathrm{m} \,. \label{eq:lensing}
\end{align}
A more general definition of these functions, including anisotropic shear stressing from neutrinos, can be constructed \citep{Bean_2010}, but we limit ourselves to the base case here. The standard equations for GR are recovered in the limits $\mu_\mathrm{mg}\rightarrow 1$ and $\Sigma_\mathrm{mg}\rightarrow 1$. In this limit, $\Psi = \Phi = \Phi_W$. As already discussed in Section \ref{sec:introduction}, in the non-linear regime many studies have been done on the effect on WL observables from a non-trivial modification to clustering (i.e. $\mu_\mathrm{mg}\neq 1$, e.g. \citealt{harnoisderaps_2022, hoyland2025fastgenerationweaklensing}), but less effort has been dedicated to the effect of the modifications to the lensing equation itself (e.g. \citealt{Barreira_2017}). As a first step, we consider only the latter and neglect the modifications to clustering, thus assuming $\mu_\mathrm{mg} = 1$ and $\Sigma_\mathrm{mg} \neq 1$. This is, of course, an approximation, which we will gradually lift in future work, implementing modified clustering in the core of FLAMINGO to get the MG density maps. Nevertheless, it shall give us an important insight into the role of modified lensing and its potential impact, complementary to what has been learned so far on the impact of modified clustering.

For simplicity, we will assume no spatial dependence and the following time dependence for $\Sigma_\mathrm{mg}$
\begin{equation}
\label{eq:Sigma}
    \Sigma_\mathrm{mg}(a) = 1 + \Sigma_0 \frac{\Omega_\Lambda(a)}{\Omega_{\Lambda, 0}}\,.
\end{equation}
The value for $\Sigma_0$ used in this work, together with other parameters used in our MG model, is listed in Table \ref{tab:MG_parameters}. This is a commonly employed parametrization (e.g. \citealt{Simpson_2012, planck_2016}) which can give a good representation of the lensing modifications on linear scales; as we approach smaller scales however, it is expected that non-linear, scale-dependent, and model-specific features of MG models start manifesting, eventually leading to a  screening of the modifications in highly dense regions. In this work, we will adopt a phenomenological description of this mechanism, implementing the screening by hand on smaller non-linear scales, as explained later in Section \ref{subsec:screening}. 
In practice, we will not solve the continuous equations of WL theory, but we will use the FLAMINGO mass shells, namely 60 discrete shells equally spaced in redshift from $z=0$ to $z=3$ (each shell has a redshift thickness of $\Delta z = 0.05$). More detailed properties of the shells are described in Section \ref{subsec:flamingo}. We need then to discretize the former equations, in particular Eqs. \eqref{eq:light_ray_position}, \eqref{eq:mag_matrix}, and \eqref{eq:lensing}. \cite{Hilbert_2009} showed that the light ray position and magnification matrix at each shell can be obtained through a recursion relation that only needs the memory from the previous two shells. At the $(n+1)$-th shell, we have \citep{Schneider_2016}
\begin{equation}
\label{eq:recursion_A}
\begin{split}
    A_{ij}^{n+1} & = \left(1 - \frac{\chi^n}{\chi^{n+1}}\frac{\chi^{n+1}-\chi^{n-1}}{\chi^n - \chi^{n-1}}\right) A_{ij}^{n-1} \\
    & + \left(\frac{\chi^n}{\chi^{n+1}}\frac{\chi^{n+1}-\chi^{n-1}}{\chi^n - \chi^{n-1}}\right)A_{ij}^n - \frac{\chi^{n+1}-\chi^n}{\chi^{n+1}}U_{ik}^n A_{kj}^n\,,
\end{split}
\end{equation}
for the magnification matrix and
\begin{equation}
\label{eq:recursion_beta}
\begin{split}
    \bm{\beta}^{n+1} & = \left(1 - \frac{\chi^n}{\chi^{n+1}}\frac{\chi^{n+1}-\chi^{n-1}}{\chi^n - \chi^{n-1}}\right) \bm{\beta}^{n-1} \\
    & + \left(\frac{\chi^n}{\chi^{n+1}}\frac{\chi^{n+1}-\chi^{n-1}}{\chi^n - \chi^{n-1}}\right)\bm{\beta}^n - \frac{\chi^{n+1}-\chi^n}{\chi^{n+1}}\bm{\alpha}^n\,,
\end{split}
\end{equation}
for the light ray position. In the previous two equations, we have introduced two new quantites, $\bm{\alpha}^n$ and $U_{ij}^n$, which are the first and second derivative of the lensing potential $\psi$, respectively. These are the deflection angle and shear matrix, respectively. The discretized lensing potential at the $n$-th shell is given by 
\begin{equation}
    \psi^n = \frac{2}{\chi^n c^2} \int_{\chi_\mathrm{min}}^{\chi_\mathrm{max}}\mathrm{d}\chi' \Phi_W\,,
\end{equation}
where $\chi_\mathrm{min}$ and $\chi_\mathrm{max}$ are the inner and outer comoving radii of the $n$-th shell, respectively, and we reinserted the speed of light. The lensing potential can be related to the mass density through the discretized 2-dimensional version of the lensing equation
\begin{equation}
\label{eq:discrete_lensing}
    K^n(\bm{\theta}) \equiv \nabla^2\psi^n = \frac{3 H_0^2\Omega_{\mathrm{m,}0}}{2c^2} \chi^n(1+z^n)\Delta\chi^n \delta^n_\mathrm{m}(\bm{\theta})\Sigma_\mathrm{mg}^n\,,
\end{equation}
where $\Sigma_\mathrm{mg}^n$ is the value of $\Sigma_\mathrm{mg}$ at the $n$-th shell, in particular evaluated at the volume weighted average redshift of the shell (the same applies to all other quantities labeled with upper $n$). The laplacian of the lensing potential at the $n$-th shell is effectively the lensing convergence at the $n$-th shell, $K^n(\bm{\theta})$, and it represents the starting point of our ray-tracing algorithm,  as explained later in Section \ref{subsec:raytracing}.


\section{Methods}
\label{sec:methods}

In this section, we present the methods used in this work. We start with a brief summary of the FLAMINGO simulation suite, then we move on to the description of the backward ray-tracing method. Finally, we show how we model screening and how we implement it in the code.

\subsection{FLAMINGO}
\label{subsec:flamingo}

As a starting point for our ray-tracing method, we use the FLAMINGO hydrodynamical simulations.  We quickly review its main features in this subsection, for a more detailed and complete description see \cite{Schaye_2023}. The FLAMINGO simulations are calibrated in order to match two observables: the galaxy stellar mass function (SMF) today and the gas mass fraction in low-z clusters of galaxies. This calibration of subgrid physics is performed via machine learning, considering observational errors and biases as well \citep{kugel2023flamingocalibratinglargecosmological}. The hydrodynamical solver is SWIFT \citep{Schaller_2024}, which is based on the smoothed particle hydrodynamics (e.g. \citealt{Price_2012}) method SPHENIX \citep{Borrow_2021}. Massive neutrinos are evolved with the particle-based $\delta f$ method \citep{Elbers_2021}. The simulations also include radiative cooling and heating rates in an element-by-element fashion \citep{Ploeckinger_2020}, time-dependent stellar mass loss by stellar winds from massive stars, asymptotic giant branch stars and supernovae \citep{Wiersma_2009}, and star formation \citep{Schaye_2007}. The implementation of energy feedback from massive stars and supernovae conserves linear momentum, angular momentum and energy and is based on the modifications of \cite{Chaikin_2023} to \cite{Dalla_Vecchia_2008}. Finally, black holes are seeded according to \cite{Di_Matteo_2005}, \cite{Bahe_2022} and their simulated physical processes are dynamical friction (using repositioning), merging and accretion \citep{Booth_2009, Bahe_2022}. Active galactic nuclei feedback is implemented with a fiducial thermal method \citep{Booth_2009}. The identification of haloes and substructures in FLAMINGO is performed using the HBT-HERONS finder \citep{moreno_2025b}.

In Table \ref{tab:simulation_parameters} we list the simulation parameters for the simulations that we use in this work. We will use both the flagship simulation, whose identifier is L2p8$\_$m9, which has a box size of 2.8 Gpc and $5040^3$ dark matter (DM) and baryonic particles, and the L1$\_$m9 simulations, which have a box size of 1 Gpc and $1800^3$ DM and baryonic particles. Every simulation also has a corresponding dark-matter-only counterpart, under the same identifier name with the suffix "$\_$DMO". The flagship simulation is run with the cosmology from DES Y3 "3×2pt+ All Ext." \footnote{This cosmology combines the DES Y3 constraints with external constraints from baryon acoustic oscillations, redshift space distortions, type Ia supernovae and Planck. The values for the parameters can be found in Table \ref{tab:cosmology_parameters}.} ("D3A", \citealt{Abbott_2022}), which we shall refer to as our fiducial cosmology, while the L1$\_$m9 simulations include different choices of cosmology (other than the fiducial one), namely the best-fit Planck cosmology ("Planck", \citealt{Planck_2020}), together with Planck cosmologies with heavier neutrinos, and one model with a lower value for $\sigma_8$ ("LS8", \citealt{Amon_2022}). We list the parameters of the fiducial cosmology, the Planck cosmology, and the low $\sigma_8$ cosmology in Table \ref{tab:cosmology_parameters}.

To obtain the WL convergence maps, we use the FLAMINGO mass maps, which are 60 concentric Healpix shells \citep{Gorski_2005} equally distributed in redshift (with thickness $\Delta z = 0.05$). The mass maps are computed with a Healpix resolution of $N_\mathrm{side} = 16384$ (corresponding to a number of pixels $N_\mathrm{pix}=12N_\mathrm{side}^2$), which translates to an angular resolution of 0.21 arcmin.\footnote{However, to make full use of Healpy, the python implementation of Healpix, we need to downsample the maps at least to the resolution $N_\mathrm{side} = 8192$.} We obtain overdensity maps for each shell $n\in\{1, \dots, 60\}$ from the mass maps using
\begin{equation}
    \delta^n(\bm{\theta}) = \frac{\Sigma^n(\bm{\theta}) - \bar{\Sigma}^n}{\bar{\Sigma}^n}\,,
\end{equation}
where $\Sigma^n(\bm{\theta})$ is the surface matter density at the $n$-th shell with average $\bar{\Sigma}^n$. This should not be confused with the modification to the lensing equation, which we call $\Sigma_\mathrm{mg}$. Note that, for the rest of this work, we will never refer to the surface matter density again, so confusion between the two quantities should not arise. Finally, since the box size is not large enough to cover the whole region of interest between redshifts $z=0$ and $z=3$, we replicate the box 5 times for the L2p8 box and 12 times for the L1 box. Moreover, we randomly rotate the replicated maps to make sure not to encounter the same structure twice. For a discussion of the rotation strategy, see Appendix A of \cite{Broxterman_2024}.
\begin{table}
    \centering
    \caption{Parameters of the FLAMINGO simulations used in this work, in their hydrodynamical and dark matter only (DMO) configurations. Columns are: name of the simulation, box size in comoving gigaparsec, number of baryonic and dark matter particles, number of massive neutrinos, initial mean baryonic particle mass, mean CDM particle mass.}
    \setlength{\tabcolsep}{4pt}
    \begin{tabular}{l|c|c|c|c|c}
    \hline
    Name & $L$ & $N_{\mathrm{b}(\mathrm{DMO})}$ & $N_\nu$ & $m_\mathrm{b}$ & $m_\mathrm{CDM}$ \\
        & [Gpc] &  &   & $[\mathrm{M}_\odot]$ & $[\mathrm{M}_\odot]$ \\
    \hline
    L1\_m9 & 1 & $1800^3$ & $1000^3$ & $1.07\times 10^9$ & $5.65\times 10^9$\\
    L1\_m9\_DMO & 1 & $1800^3$ & $1000^3$ & $0$ & $6.72\times 10^9$\\
    L2p8\_m9 & 2.8 & $5040^3$ & $2800^3$ & $1.07\times 10^9$ & $5.65\times 10^9$ \\
    L2p8\_m9\_DMO & 2.8 & $5040^3$ & $2800^3$ & $0$ & $6.72\times 10^9$ \\
    \hline
    \end{tabular}
    \label{tab:simulation_parameters}
\end{table}

\begin{table*}
    \centering
    \caption{The three cosmologies in FLAMINGO that we will use in this work: the fiducial cosmology (DES Y3 "3×2pt+ All Ext."), the best-fitting Planck cosmology and a low $\sigma_8$ cosmology. From left to right we have the name of the cosmology, the reduced Hubble constant $h = H_0 / (100\mathrm{km/s/Mpc})$, the present-day density parameters for total matter, baryonic matter, cosmological constant and neutrinos, the amplitude and spectral index of the primordial power spectrum and the amplitude of the linear matter power spectrum today parameterized as the r.m.s. mass density fluctuation in spheres of radius $8 h^{-1}\mathrm{Mpc}$.} 
    \begin{tabular}{c|c|c|c|c|c|c|c|c}
    \hline
    Name & $h$ & $\Omega_\mathrm{m}$ & $\Omega_\mathrm{b}$ & $\Omega_\Lambda$ & $\Omega_\nu$ & $A_\mathrm{s}$ & $n_\mathrm{s}$ & $\sigma_8$ \\
    \hline
    D3A & 0.681 & 0.306 & 0.0486 & 0.694 & 1.39$\times 10^{-3}$ & 2.099$\times 10^{-9}$ & 0.967 & 0.807 \\
    Planck & 0.673 & 0.316 & 0.0494 & 0.684 & 1.42$\times 10^{-3}$ & 2.101$\times 10^{-9}$ & 0.966 & 0.812 \\
    LS8 & 0.682 & 0.305 & 0.0473 & 0.695 & 1.39$\times 10^{-3}$ & 1.836$\times 10^{-9}$ & 0.965 & 0.760 \\
    \hline
    \end{tabular}
    \label{tab:cosmology_parameters}
\end{table*}
\subsection{Ray-tracing method}
\label{subsec:raytracing}

\begin{figure*}
  \centering
  \includegraphics[width=\textwidth]{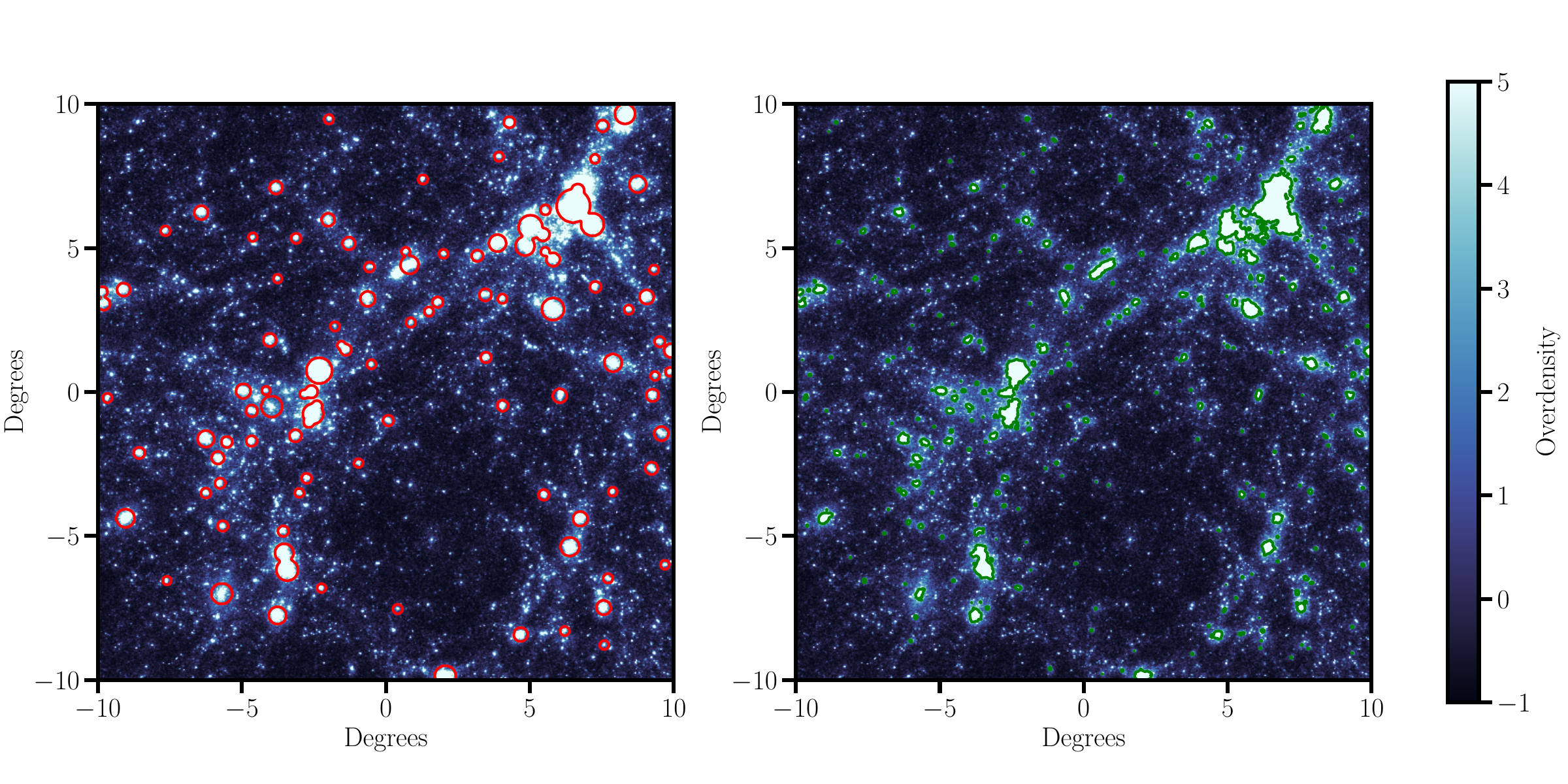}

  \caption{We plot a 20x20 degrees portion of the overdensity map in the first shell to illustrate our two screening methods. The pixels inside the colored contours are screened, thus undergoing regular GR lensing, while the regions outside have MG lensing. Left: screening is performed inside spherical shells of radius $R_\mathrm{200c}$ centered on haloes of mass $M_{200c} > M_\mathrm{thresh} = 10^{13}\mathrm{M}_\odot$ (red contours). Right: screening is performed in regions of density higher than $\delta_\mathrm{thresh}=5$ (green contours).}
  \label{fig:overdensity_and_hitmaps}
\end{figure*}

In this subsection, we summarize the ray-tracing algorithm presented in \cite{Broxterman_2024}, and explain how it is modified to include modified gravity lensing \footnote{The base version of the code can be found at \url{https://github.com/JegerBroxterman/Lensing_raytrace_FLAMINGO}.}. We begin the ray-tracing algorithm by aiming a light ray at the center of each pixel in the innermost shell (the closest to the observer). For the subsequent shells, we solve the lensing equation in harmonic space, where it becomes analytical
\begin{equation}
    \psi^n_{\ell m} = \frac{-2K^n_{\ell m}}{\ell(\ell+1)}\,.
\end{equation}
Here, $K^n_{\ell m}$ is simply the harmonic space transform of the right-hand side of Eq. \eqref{eq:discrete_lensing}, which can be easily obtained for each shell, as it contains only constants, geometrical quantities, and the input overdensity map. We then compute the deflection angle and shear matrix harmonic coefficients $\alpha^n_{\ell, m}$, $U^n_{\ell, m, (i,j)}$ as the first and second derivatives of the lensing potential $\psi^n_{\ell, m}$\footnote{The derivatives are computed using Healpy.}, and we convert them back to real space. Using the recursion relations Eqs. \eqref{eq:recursion_A}, \eqref{eq:recursion_beta} we find the angular position of each light ray $\bm{\beta}$ and the magnification matrix $A_{ij}$, from which we extract the convergence through Eq. \eqref{eq:convergence_from_A}. Finally, we add this contribution to the total convergence via the discretized version of Eq. \eqref{eq:convergence_integrated}, i.e.
\begin{equation}
    \kappa^{n+1} = \kappa^{n} + \left(1 - \frac{A^{n+1}_{11} + A^{n+1}_{22}}{2}\right)n^{n+1}\Delta\chi^{n+1}\,,
\end{equation}
where $n^{n+1}$ and $\Delta\chi^{n+1}$ are the source distribution (see Eq. \eqref{eq:source_distribution} below) and the thickness of the $(n+1)$th shell, respectively. Repeating the process for each shell yields the final convergence map as an integrated 2-dimensional map on the sphere, $\kappa(\bm{\theta})$. Note that, in general, the light ray positions will not be aimed at the pixel centers in any shell different from the first one (where they are by construction). In order to make use of the built-in functions of Healpy, that can compute the derivatives of the lensing potential at each pixel centre, we use a nearest grid point (NGP) method to assign each light ray position to a pixel (namely, the pixel closest to the light ray position), and we parallel transport the quantities computed by Healpy at the NGP pixel to the light ray position using the method of \cite{becker2012calclensweaklensingsimulations}, so that all quantities in the recursion relations are evaluated at the same point and can be summed together.

We choose to employ a \textit{Euclid-like} source distribution \citep{Euclid_2020} in the integrated convergence
\begin{equation}
\label{eq:source_distribution}
    n(z) \propto \left(\frac{z}{z_0}\right)^2\exp\left[-\left(\frac{z}{z_0}\right)^{3/2}\right]\,,
\end{equation}
where $z_0 = 0.9/\sqrt{2}$. The source density is normalized to unity $\int n(z)\mathrm{d}z = 1$ and satisfies $n(z)\mathrm{d}z = n(\chi) \mathrm{d}\chi$. More general source distributions can easily be implemented, but we will not focus on exploring different ones in this work. 

We modify the ray-tracing algorithm just presented to include the possibility of having the lensing equation solved in modified gravity theories. The main structure of the code remains unchanged, and our difference with respect to \cite{Broxterman_2024} is that we include a $\Sigma^n_\mathrm{mg}$ different from 1 (which would be the standard GR lensing case) in solving Eq. \eqref{eq:discrete_lensing}. We explain how we construct the $\Sigma^n_\mathrm{mg}$ function to include both time dependence and screening in the next subsection.

\subsection{Screening}
\label{subsec:screening}

MG models need to encode a successful screening mechanism (for reviews, see e.g. \citealt{Joyce_2015} and \citealt{Koyama_2016}) ensuring that GR is restored in high-density regions or near massive objects. Within the broad landscape of MG models that introduce a scalar field, this mechanism arises from nonlinear terms in the scalar field equation of motion. Although there are four classes of screening, the specific mechanism and its response to local density and compact sources generally depend on the particulars of a given MG model. Some examples include a coupling to matter of the fifth force (the one sourced by the additional scalar degree of freedom of the MG model) that is dependent on the environment, and that becomes small in regions of high density \citep{Hinterbichler_2010}. Other possibilities encode the screening through the scalar field itself acquiring a large mass near compact objects, making the fifth force effects short ranged and unobservable \citep{Khoury_2004}. However, since we are not testing a single specific MG model, but rather exploring the influence of MG on ray-tracing, we do not implement a specific  screening mechanism explicitly in the functional form of $\Sigma_\mathrm{mg}$, but we instead screen "by hand". We will model screening in two different ways, one based on halo positions and another one based on over-dense regions, to represent the phenomenology described above. We will call these the halo method and the overdensity method, respectively. 

The first method, exploits the halo data that is accessible in the L1\_m9 hydrodynamical runs in FLAMINGO. This allows us to find the position of the center of each halo in the simulation (angular position and redshift), we can then link  them to Spherical Overdensity APerture (SOAP) catalogues \citep{McGibbon_2025}, giving us $R_{200c}$ and $M_{200c}$, which we will refer to as virial radius and virial mass, respectively. These are defined as the radius within which the average density of the halo is 200 times the critical density of the Universe, and the corresponding mass.
By selecting a mass threshold $M_\mathrm{thresh}$, our algorithm will match the positions of all haloes of mass $M_{200c} > M_\mathrm{thresh}$ to the corresponding Healpix pixel, and for each of such haloes, it will also find all the pixels that lie within a sphere of radius $R_{200c}$ centered on the halo center pixel. By doing so, we effectively set the screening radius to coincide with the virial radius of the source. In the next section, we will explore different choices for the screening radius. Now, taking the overdensity map at the $n$-th shell, in each of the selected pixels from the above procedure (call them "halo$\_$pixels"), we set $\Sigma_\mathrm{mg}^n = 1$, therefore leaving the overdensity effectively screened and thus unchanged, while we multiply all remaining pixels by the value of $\Sigma_\mathrm{mg}^n$ given by Equation \eqref{eq:Sigma}. In this way, we construct a new proxy density map $\delta'^n$ at each shell such that the value in each pixel is $\delta'^n_\mathrm{pix} = \delta_\mathrm{pix}^n \Sigma^n_\mathrm{pix}$, where
\begin{equation}
    \Sigma^n_\mathrm{mg, pix} =
    \begin{cases}
        1 \quad \quad\;\;\mathrm{if\; pix\in halo\_pixels} \\
        \Sigma^n_\mathrm{mg} \quad \;\mathrm{otherwise}
    \end{cases}
\end{equation}
This creates sharp discontinuities in the density maps, but we have checked that a smoother transition between the screened and unscreened regions does not affect the results. We will lift this simplified assumption and study screening more rigorously in future work. We can finally use this new overdensity map as the input of the standard ray-tracing method described before.

In the left panel of Figure \ref{fig:overdensity_and_hitmaps} we show an example of what this screening procedure looks like. We plot a 20x20 degree region of the first overdensity shell ($z\approx0.05$), together with red contours where the MG theory is screened back to GR. These circles are centered on haloes of mass higher than a given threshold ($M_\mathrm{thresh} = 10^{13} \mathrm{M}_\odot$ in the example plot) and have a radius equal to the virial radius $R_{200c}$ of the halo. Non-circular regions indicate parts of the sky where multiple haloes are close to each other. The ray-tracing algorithm is applied as standard GR lensing in the pixels that lie inside the red circles and as MG lensing in the pixels that lie outside those circles. It can be seen from the figure that the halo positions are, on average, well matched with high overdensity regions. On the other hand, there are some portions of the density map where high overdensities are not contained in the red circles. For this reason, we also implement a second screening mechanism, which instead screens the MG model where the overdensity value exceeds a given threshold, irrespective of the presence of haloes. This means that the "halo$\_$pixels" set is now the set of all pixels where $\delta_\mathrm{pix} > \delta_\mathrm{thresh}$, where $\delta_\mathrm{thresh}$ is the selected overdensity threshold.

We show this different approach in the same region of the sky in the right panel of Figure \ref{fig:overdensity_and_hitmaps}, where we add green contours within which the overdensity is higher than the threshold $\delta_\mathrm{thresh}=5$. Note that, as discussed above, there are regions that lie within the green contours but not within the red contours. This will be reflected in the power spectrum of the integrated convergence map, as we will show later. Note that this behavior will be present in each shell, and that a given angular direction on the sky can go from screened to unscreened and back as we follow the light ray path through the redshift shells.

Both prescriptions can be viewed as phenomenological realizations of known screening behaviors. The halo-based method mimics Vainshtein-like screening, where non-linear derivative interactions suppress the fifth force in the vicinity of massive objects \citep{Nicolis_2009}, while the overdensity-based method is more reminiscent of chameleon-like screening, in which the effective mass of the scalar field increases with the local matter density \citep{Khoury_2004}. In practice, screening mechanisms in realistic MG models depend on the full environmental density field rather than exclusively on halo locations. For this reason, we expect the overdensity-based method to provide a more realistic representation of the behavior of viable screening scenarios, although both approaches allow us to explore possible MG phenomenologies and to assess the impact of screening on the lensing signal.

\begin{table}
    \centering
    \caption{Parameters of our modified gravity model. From left to right: the discrepancy of $\Sigma_\mathrm{mg}$ from $1$ today (defined in Eq. \ref{eq:Sigma}), the dark energy equation of state parameters in the Chevallier-Polarski-Linder  parametrization $w(a) = w_0 +w_a(1-a)$, the mass and density thresholds for the two screening methods.}
    \setlength{\tabcolsep}{4pt}
    \begin{tabular}{c|c|c|c|c}
    \hline
    $\Sigma_0$ & $w_0$ & $w_a$ & $M_\mathrm{thresh}$ & $\delta_\mathrm{thresh}$ \\
    \hline
    0.044 & -0.727 & -1.05 & $10^{13} \mathrm{M}_\odot$ & 5 \\
    \hline
    \end{tabular}
    \label{tab:MG_parameters}
\end{table}


\section{Results and discussion}
\label{sec:results}

\begin{figure}
  \centering
  \includegraphics[width=\columnwidth]{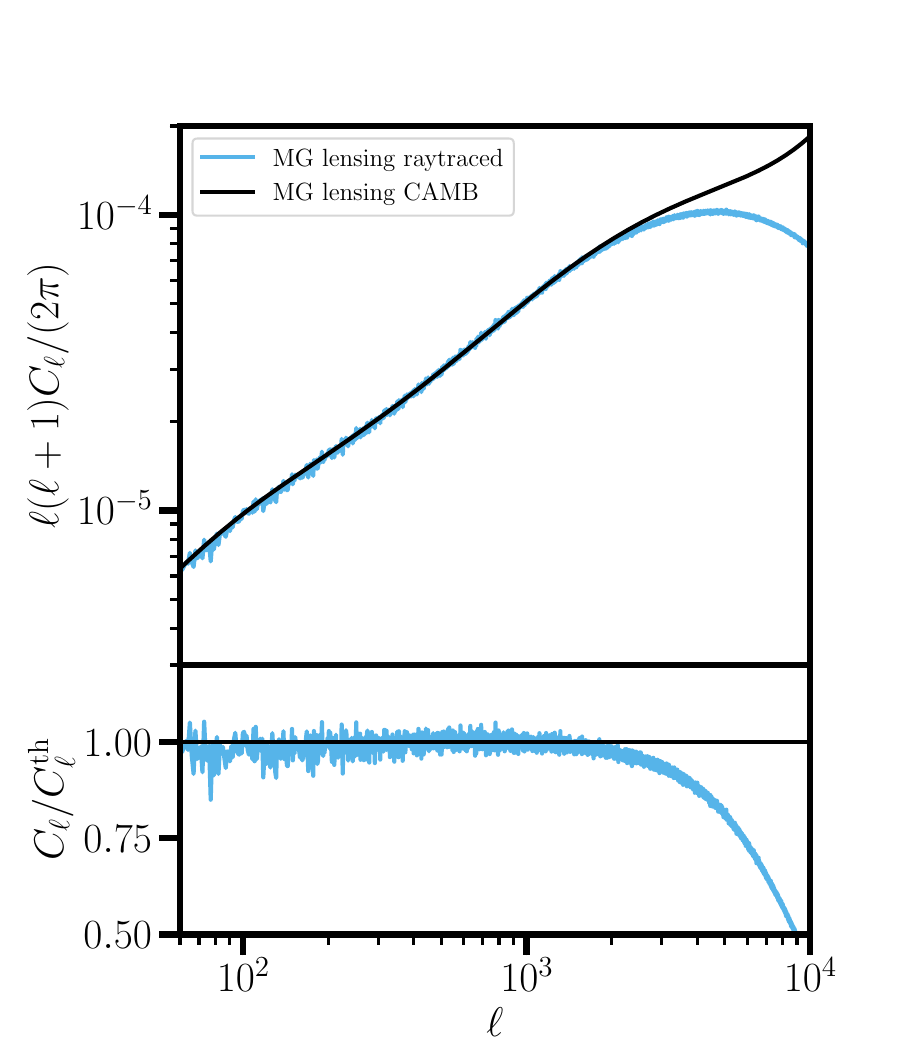}

  \caption{Top: angular power spectra of the MG weak lensing convergence obtained from CAMB (black line) and obtained from the ray-tracing algorithm (cyan line) with a purely time dependent $\Sigma(a)$. Bottom: ratio to the CAMB power spectrum. The resolution of the density maps in the ray-traced case is $N_\mathrm{side}=4096$. The small scale suppression is due to the pixelization of the Healpix map.}
  \label{fig:cls_theory_MG}
\end{figure}

We begin the presentation of our results by comparing the ray-traced convergence power spectrum obtained with a time dependent only $\Sigma_\mathrm{mg}$ (i.e. without screening) with the theoretical prediction from CAMB \citep{Lewis_2000}. The ray-traced spectrum is obtained as the average of the 8 observers for which the lightcone output is stored in the L2p8\_m9 simulation. We compute the theoretical angular power spectrum using the extended Limber approximation \citep{Limber:1954zz, Loverde_2008}
\begin{equation}
\label{eq:limber}
    C_\ell = \int_0^{\chi_\mathrm{fin}} \mathrm{d}\chi \frac{W^2(\chi)}{\chi^2} P\left(\frac{\ell + 1/2}{\chi}, z(\chi)\right)\,,
\end{equation}
where $\chi_\mathrm{fin}$ is the endpoint of the ray-tracing (in our case $\chi_\mathrm{fin} = \chi(z=3)$), $P$ is the matter power spectrum and $W$ is the weak lensing kernel, given by \citep{Kaiser:1991qi}
\begin{equation}
\label{eq:kaiser}
    W(\chi) = \frac{3H_0^2\Omega_{m,0}}{2c^2}\chi \left[1+z(\chi)\right]\Sigma_\mathrm{mg}(\chi)\int_\chi^{\chi_\mathrm{fin}}\mathrm{d}\chi' n(\chi') \frac{\chi' - \chi}{\chi'}\,,
\end{equation}
where we have added the $\Sigma_\mathrm{mg}(\chi)$ to take MG into account. All the cosmological quantities are computed using CAMB. The non-linear part of the matter power spectrum is obtained through the Mead2020 method without baryon correction \citep{Mead_2021}. We show the comparison between the ray-traced spectrum and the theoretical one in Figure \ref{fig:cls_theory_MG}. We note that the two spectra are well in agreement up to a scale of $\ell$ between $1000$ and $2000$, similarly to what happens in the GR case \citep{Broxterman_2024} for the chosen resolution of $N_\mathrm{side}=4096$. The small scale difference is due to the pixelization of the Healpix map.

\begin{figure}
  \centering
  \includegraphics[width=\columnwidth]{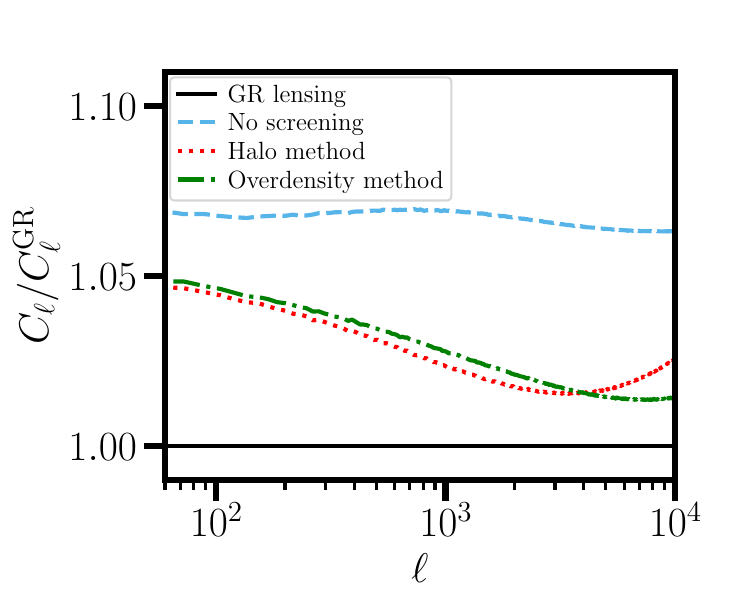}

  \caption{Ratios of the angular power spectra of the integrated convergence map with respect to the standard GR lensing prescription (continuous black line). The dashed cyan line corresponds to MG lensing without screening, while the dotted red and dot dashed green lines correspond to MG lensing with screening implemented using the halo method (with $M_\mathrm{thresh}=10^{13}\mathrm{M}_\odot$) and using the overdensity method (with $\delta_\mathrm{thresh}=5$) respectively. For all three MG lines $\Sigma_0=0.044$.}
  \label{fig:cls_delta_vs_haloes}
\end{figure}

We now proceed with the inclusion of our two MG lensing prescriptions with the screening methods that we described earlier, and we will compare them to the standard GR lensing case. In Figure \ref{fig:cls_delta_vs_haloes}, we show the power spectra of the integrated convergence map obtained with MG lensing using the two different screening mechanisms presented in Section \ref{subsec:screening}, and we compare them with the standard GR lensing one. In particular, we only plot the ratio of the various MG spectra with respect to the GR lensing one, as the power spectra would all resemble the top plot of Figure \ref{fig:cls_theory_MG}. Note that, with respect to Figure \ref{fig:cls_theory_MG}, we have also smoothed the ray-traced lines for clarity, to avoid confusion in the plots. The smoothing is applied only to the plotted curves by averaging the $C_\ell$ values over bins of width $2^4$ in $\ell$; the underlying convergence maps and power spectra remain unchanged. This will not affect the physical interpretation of our results, and we will do the same in future plots. All spectra have been obtained using the L1\_m9 hydrodynamical runs of FLAMINGO (since the halo positions in the lightcones are only available there). The thresholds chosen for the two methods are the same as used in Figure \ref{fig:overdensity_and_hitmaps}, i.e., $M_\mathrm{thresh} = 10^{13} \mathrm{M}_\odot$ for the halo based method and $\delta_\mathrm{thresh} = 5$ for the overdensity based method. We also include a non-screened line with MG lensing to show the maximum difference we could expect between the two regimes. For all MG lines, we have chosen $\Sigma_0 = 0.044$ (best fit value from \citealt{DESI_2024VII}, "DESI + CMB + DESY3"). Although DESI cannot constrain $\Sigma_0$ well by itself, we chose the DESI value to be consistent with the choices of $w_0$ and $w_a$. The value is very similar to the one from DESY3 \citep{Abbott_2023}, as is expected since DES is considered as an additional probe when obtaining $\Sigma_0$ in DESI. Later, we will show the impact of varying this parameter. We find up to a 5\% discrepancy between GR and MG lensing with both screening methods at the largest scales ($\ell\approx100$), which decreases to smaller values at smaller scales (higher $\ell$), as expected from screening.

Figure \ref{fig:cls_delta_vs_haloes} shows that the behavior of the two screening methods is similar on larger scales, but differs on smaller scales, with the halo based method showing a dip and regain of power towards the end of the spectrum. As anticipated in Section \ref{sec:methods}, this difference at small scales is due to the fact that some of the regions of high overdensity are not captured inside a screening radius distance from the center of the haloes (see Fig. \ref{fig:overdensity_and_hitmaps}). A region of high overdensity will be relevant in the lensing equation Eq. \eqref{eq:lensing} (it will have a non negligible influence on the deflection of the light ray), and if that region is not screened in the MG lensing, then it will enter the modified lensing equation (i.e. with a value of $\Sigma_\mathrm{mg}$ different from one). In other words, screening with spheres of a certain radius will imprint in the power spectrum of the convergence a characteristic scale at which screening is at its peak efficiency. We note that at large scales, the screened lines do not converge to the same value as the unscreened one, contrarily to what one would expect considering that screening activates on small, non-linear scales. This is a consequence of our phenomenological implementation of screening. Despite being only a small fraction of the total volume (typically a few percent, depending on the density threshold and the shell redshift), the screened regions are a biased selection of the whole sky, located where the overdensity is highest. As these regions account for a significant fraction of the lensing signal, screening the MG contribution within them affects the convergence power spectrum on all scales. Our screening prescription will then not only imprint a scale-dependent feature in the power spectrum, but also lower its amplitude on the largest scales. Therefore, within this implementation there is no one-to-one correspondence between screened (unscreened) regions in the overdensity field and small (large) scales in the convergence power spectrum. We stress nevertheless, that the framework works fully for models that have a scale-independent $\Sigma_\mathrm{mg}$, such as Generalized-Brans-Dicke theories (e.g. \citealt{Pogosian_2016}).

\begin{figure}
  \centering
  \includegraphics[width=\columnwidth]{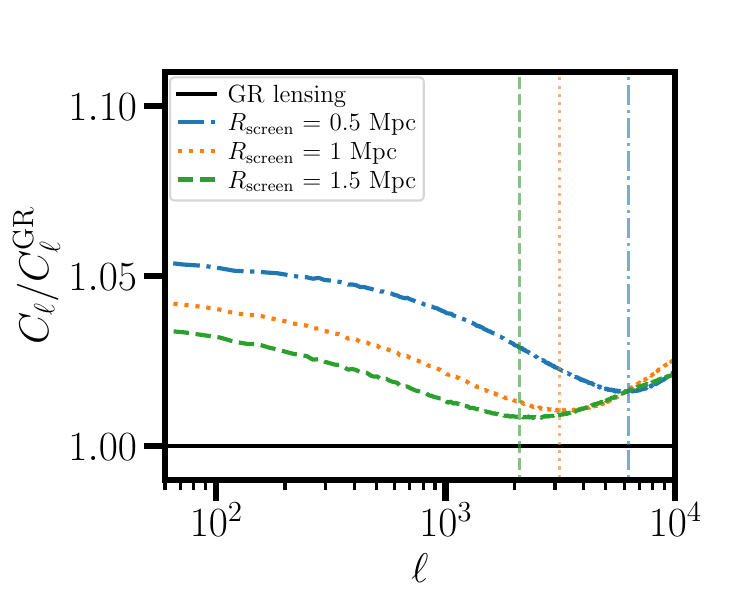}

  \caption{Ratio of the convergence power spectrum with source distribution as $n(z) = \delta_\mathrm{D}(z=1)$ with respect to the standard GR convergence spectrum. Regular GR lensing is in continuous black, while the colored lines are MG lensing with the same screening radii for all haloes and equal to 0.5 Mpc, 1 Mpc and 1.5 Mpc respectively for dot dashed blue, dotted orange and dashed green. Vertical lines with corresponding color and style indicate the approximate scale that we expect to be imprinted from our screening procedure at the distance where lensing is most efficient (half of the distance where the sources are located).}
  \label{fig:ratios_source_delta}
\end{figure}

Let  us investigate the impact of the specific choices we make in  each of the two approaches  to screening. We start with the halo method and explore different choices for the screening radius, setting it to a constant instead of considering the specific virial radius $R_{200c}$ of each halo. In order to have a clearer picture of the ray-traced spectra for these cases, we make the simplification of distributing  the light ray sources  as a Dirac delta, $\delta_\mathrm{D}$, peaked at redshift $z=1$ instead of the more complicated \textit{Euclid-like} distribution. The lensing will then have its most relevance at a distance equal to half the distance where the sources are located. By making this assumption, we expect to imprint a feature in the power spectrum at the scale corresponding to the angular scale under which we see the halo spheres at this distance. This means that we expect the dip of power to happen at a value of $\ell$ given approximately by
\begin{equation}
    \ell_\mathrm{screen} \approx \frac{\pi}{\theta_\mathrm{screen}} \approx \pi\frac{d_A}{2R_\mathrm{screen}}\,,
\end{equation}
where $\theta_\mathrm{screen}$ is the angle under which we see the halo screening sphere and $d_A$ is the angular diameter distance. 
In Figure \ref{fig:ratios_source_delta}, we show  the  behaviour of the $C_\ell$'s for $R_{\rm screen}=0.5, 1, 1.5 \;\mathrm{Mpc}$, which are representative of the range of values that $R_{200}$ takes for the haloes of mass $M\approx10^{13}\mathrm{M}_\odot$. We add a dashed vertical line in the  plot corresponding to the value of $\ell_\mathrm{screen}$ for each case, and we see that the dip occurs around this value, confirming our interpretation of the dip in Figure \ref{fig:cls_delta_vs_haloes}. In the most general case, the dip will also be affected by the complete source distribution and by the different values of the halo screening radii, so we will not be able to determine a precise value for $\ell_\mathrm{screen}$, but this is where it originates from.

\begin{figure}
  \centering
  \includegraphics[width=\columnwidth]{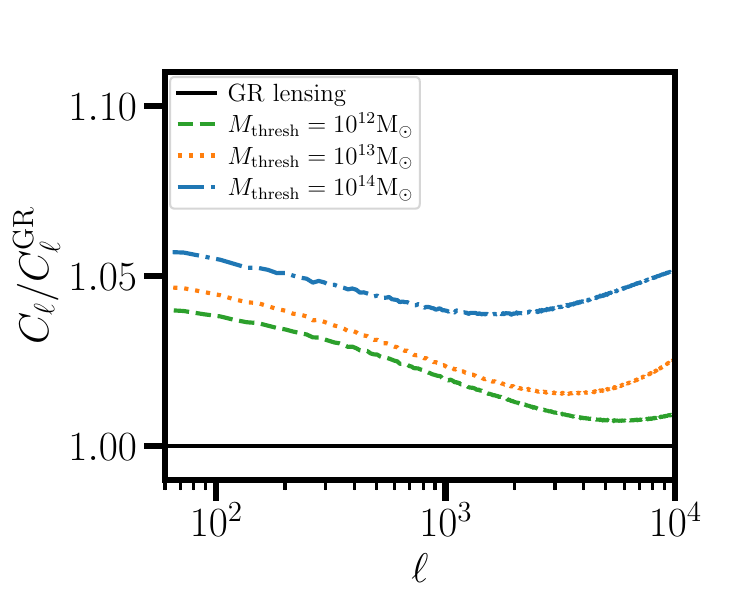}

  \caption{Ratio of the convergence power spectrum obtained via MG lensing with the halo based method with respect to the standard GR lensing spectrum. Different colors and line styles represent different choices for the mass threshold $M_\mathrm{thresh}$.}
  \label{fig:cls_ratios_screening_haloes}
\end{figure}

\begin{figure}
  \centering
  \includegraphics[width=\columnwidth]{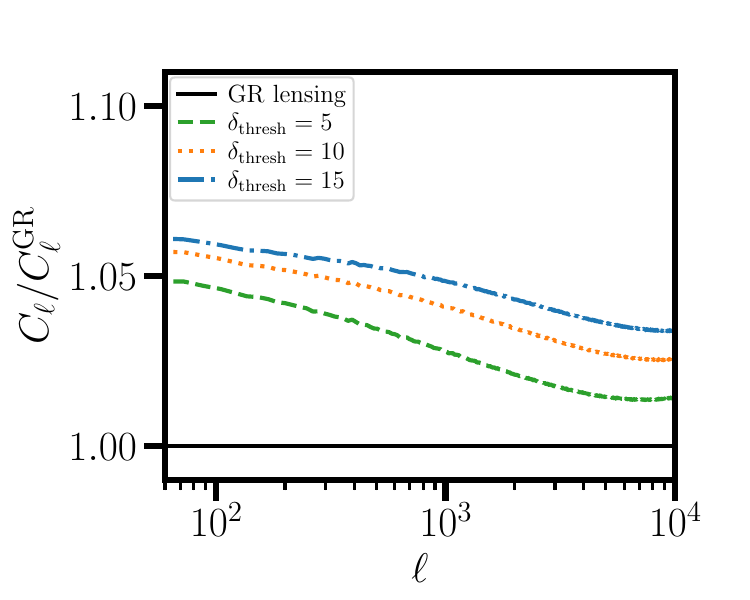}

  \caption{Ratio of the convergence power spectrum obtained via MG lensing with the overdensity based method with respect to the standard GR lensing spectrum. Different colors and line styles represent different choices of the overdensity threshold $\delta_\mathrm{thresh}$.}
  \label{fig:cls_ratios_screening_delta}
\end{figure}

Armed with an understanding of the features, we now return to the more realistic scenario with the \textit{Euclid-like} source distribution and explore the impact of varying our MG parametrization, starting with the screening thresholds. In Figures \ref{fig:cls_ratios_screening_haloes} and \ref{fig:cls_ratios_screening_delta}, we show the effect of varying the mass threshold and overdensity threshold in our two screening methods. In particular, we show the ratio between the MG lensing power spectra with the GR lensing one, for three different values of the mass and three different values of  overdensity. We see that increasing the threshold leads to less screened power spectra. This is expected: in both cases we screen all pixels that are above the threshold, so increasing the threshold means decreasing the number of pixels that will be screened, leading to power spectra that will  differ more from the GR one. Note that the different behavior of the two methods, already addressed in Figure \ref{fig:cls_delta_vs_haloes}, is also reflected when varying the thresholds: the halo method always shows a dip and regain of power, while that is not the case for the overdensity method.

\begin{figure}
  \centering
  \includegraphics[width=\columnwidth]{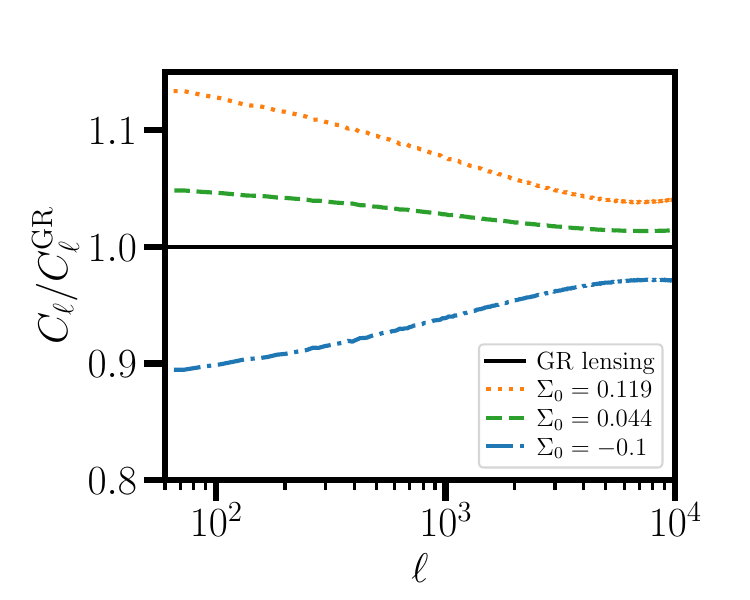}

  \caption{Ratio of the convergence power spectrum obtained via MG lensing with the overdensity based method with respect to the standard GR lensing. Different colors and line styles represent different choices of the MG parameter $\Sigma_0$.}
  \label{fig:cls_sigma0}
\end{figure}

In Figure \ref{fig:cls_sigma0}, we show the effect of varying the MG parameter $\Sigma_0$, defined in Equation \ref{eq:Sigma}. Note that the range on the y-axis is different from previous figures to make all the lines visible and to compare with future plots. We took two best-fitting values from the analysis in \cite{DESI_2024VII}, namely the "DESI+CMB" one ($\Sigma_0=0.119$) and the "DESI+CMB+DESY3" one ($\Sigma_0=0.044$). Moreover, we also show a negative value, despite current constraints from DESI favoring $\Sigma_0>0$, to explore the full range of possibilities. Considering a negative value will also be helpful later, when we will compare the MG model choices with cosmology variations. In all three cases, we used the overdensity screening method with $\delta_\mathrm{thresh}=5$. We see that the discrepancy between the MG lensing power spectrum and the GR lensing one at the largest scales is heavily affected by the choice of $\Sigma_0$. The sign of this parameter determines if the $C_\ell$ for the chosen MG model has a higher or lower amplitude with respect to the GR one, while its absolute value is related to the magnitude of the discrepancy.

\begin{figure}
  \centering
  \includegraphics[width=\columnwidth]{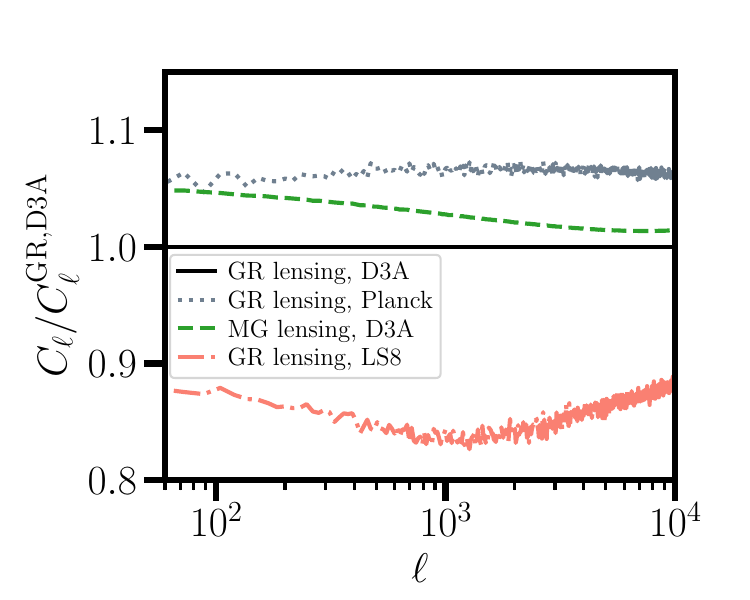}

  \caption{Comparison of the effect on the convergence power spectrum of changing the cosmology with respect to changing the lensing prescription. The lines are obtained as the ratio of the power spectrum with the standard GR lensing, fiducial cosmology spectrum (continuous black): the GR lensing Planck cosmology in dotted silver, the GR lensing low $\sigma_8$ cosmology in dot dashed pink and the MG lensing (with overdensity screening method), fiducial cosmology in dashed green.}
  \label{fig:cls_cosmo_vs_MG}
\end{figure}

We proceed in our analysis by comparing the effects of our changes with those of other physical mechanisms that generate variation in the WL convergence power spectrum. We will explore the relative impact of changes in the cosmological parameters and of baryonic feedback associated with galaxy formation. In Figure \ref{fig:cls_cosmo_vs_MG}, we show a comparison between varying the cosmology in a GR lensing scenario with respect to keeping the cosmology fixed but switching from the GR lensing prescription to the MG lensing one. The MG spectrum is the same as that shown in Figure \ref{fig:cls_delta_vs_haloes}, obtained with the overdensity method. As already mentioned in Section \ref{subsec:flamingo}, the simulations with different cosmologies are available only in the (relatively) smaller box of size 1 Gpc. In the plot, we show GR lensing lines obtained with the fiducial, the Planck, and the low-$\sigma_8$ cosmologies, and we also show the MG lensing line with the fiducial cosmology (using the overdensity screening method with $\delta_\mathrm{thresh}=5$ and $\Sigma_0=0.044$). We find that the Planck cosmology gives rise to a similar overall amplitude shift in the power spectrum of the convergence as the MG spectrum - around $5\%$ - at the largest scales, and this is due to our choice of $\Sigma_0$, as we saw in Figure \ref{fig:cls_sigma0}. However, the Planck line differs in behavior going towards smaller scales, where it remains a constant shift with respect to the standard line, whereas the MG line gets screened. The low $\sigma_8$ line results in a lower amplitude of the power spectrum with respect to the fiducial cosmology, which is not matched by the MG models with a positive value for $\Sigma_0$, and is more similar to the $\Sigma_0=-0.1$ model. Since current constraints on $\Sigma_0$ seem to slightly favor $\Sigma_0>0$ (still consistent with the GR value, see e.g. \citealt{Abbott_2023, DESI_2024VII}), we conclude that the cosmology with lower $\sigma_8$ is not degenerate with MG lensing effects, as that would require $\Sigma_0<0$. We will show this more quantitatively in a later plot. We conclude that cosmology variations can be as relevant as MG lensing effects at the largest scales, depending on the parameters of the MG model and on the values of the cosmological parameters.

\begin{figure}
  \centering
  \includegraphics[width=\columnwidth]{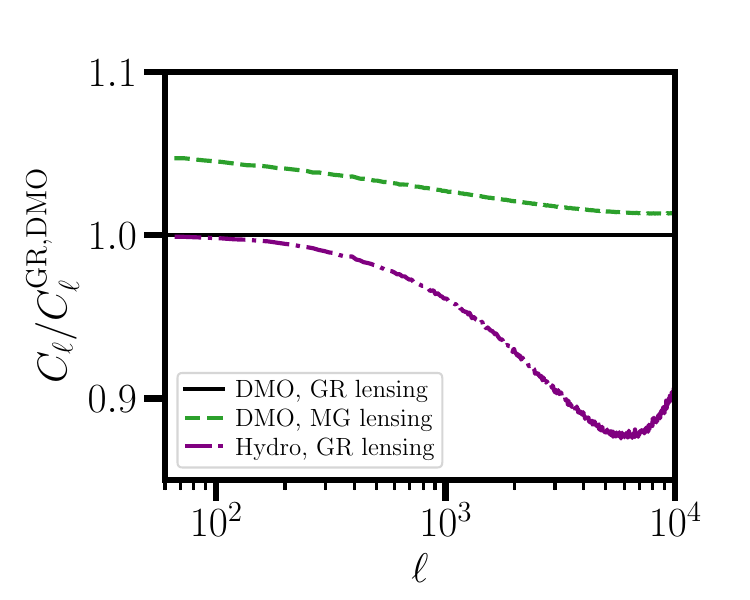}

  \caption{Comparison of the effect on the convergence power spectrum of baryonic feedback with respect to changing the lensing prescription. The lines are obtained as the ratio of the power spectrum with the standard GR lensing DMO spectrum. We show the standard line in continuous black, the GR lensing hydrodynamical spectrum in dot dashed purple and the MG lensing DMO spectrum (with overdensity-based screening) in dashed green.}
  \label{fig:cls_hydro_vs_MG}
\end{figure}

In Figure \ref{fig:cls_hydro_vs_MG}, we show the comparison between a spectrum obtained from a simulation with baryon modeling via the regular GR lensing prescription and a DMO spectrum obtained via the MG lensing prescription, both relative to the standard DMO GR case\footnote{See \cite{Schaller_2025} for a detailed discussion of the baryonic effects on the matter power spectrum in the FLAMINGO model.}. We see that the hydrodynamical case agrees with the DMO one at low $\ell$ and deviates from it at higher $\ell$, in opposition to the behavior of the MG lensing spectrum. The maximum discrepancy between the hydrodynamical spectrum and the DMO one is reached around $\ell \approx 5000$ and is about $15\%$. This is greater than the maximum discrepancy between the MG lensing spectrum and the GR lensing one (which is about $5\%$ as already discussed), although in this case the peak happens at the largest scales, in the low $\ell$ part of the spectrum. The effect that is most important to take into account then depends on the scales that are analyzed. In the next paragraph, we continue to compare the two types of effects.

\begin{figure}
  \centering
  \includegraphics[width=\columnwidth]{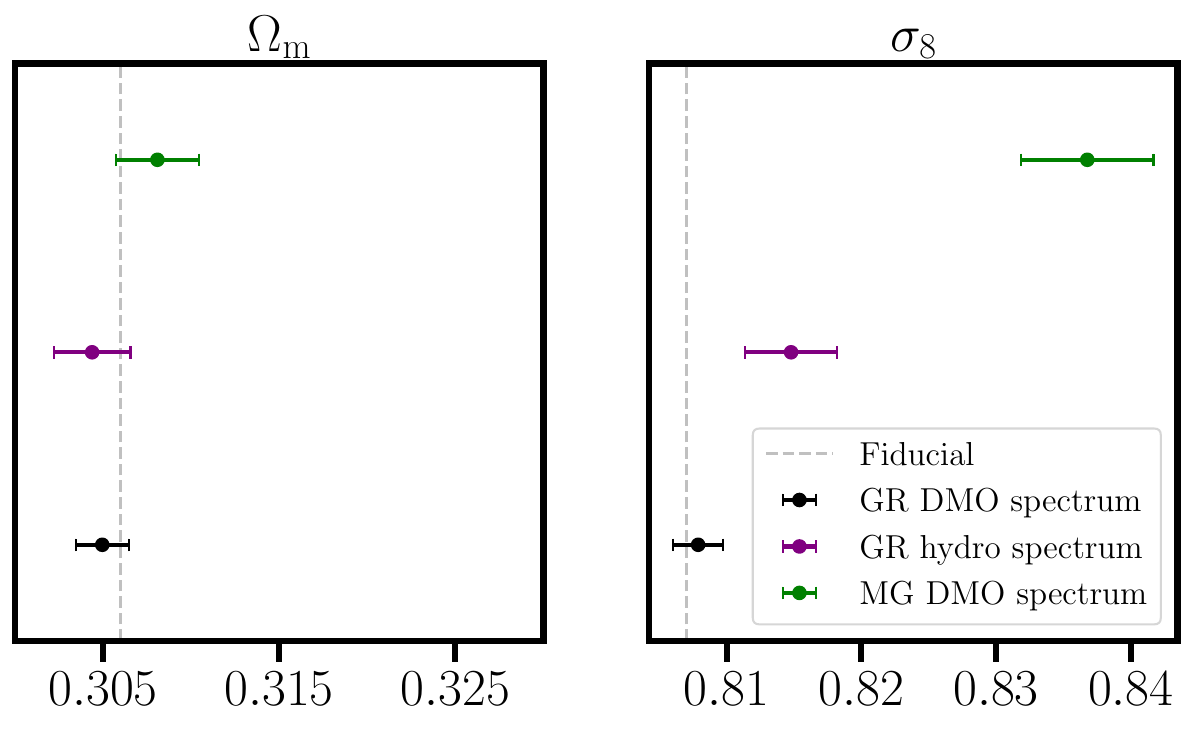}

  \caption{Best-fitting values of the cosmological parameters $\Omega_\mathrm{m}$ (left) and $\sigma_8$ (right) when fitting the convergence power spectrum with a standard GR DMO \LCDM{} model. The black points are fit to a GR DMO spectrum, the purple points to a hydrodynamical GR spectrum and the green points to a MG DMO spectrum (with $\Sigma_0=0.044$). All three spectra have been produced with the fiducial D3A cosmology, whose parameter values are shown as grey dashed vertical lines.}
  \label{fig:best_fit_parameters}
\end{figure}

As a final step of this work, we want to get a more quantitative estimate of the relevance of MG effects on the WL convergence power spectrum. Thus, we investigate the bias in selected \LCDM{} parameters that arises when fitting a MG lensing spectrum with a \LCDM{} model, relative to the results obtained when the same analysis is applied to GR lensing spectrum. We will compare it to the bias we would get if we instead fit a hydrodynamical spectrum with a DMO model as a reference case for a well-known level of systematics (e.g. \citealt{Semboloni_2011, Gouin_2019}). To do so, we treat the power spectrum that we extract from our ray-tracing algorithm as the data to be fit. The estimate of the error bar on the data is given by
\begin{equation}
    \sigma_\ell = \sqrt{\frac{2}{(2\ell + 1)f_\mathrm{sky}}} \left(C_\ell + N\right)\,,
\end{equation}
where $f_\mathrm{sky}$ is the sky fraction covered by the survey of reference and $N$ is given by
\begin{equation}
    N = \frac{\sigma_\epsilon^2}{\bar{n}}\,.
\end{equation}
In the above expression, $\sigma_\epsilon^2$ is the variance of the observed ellipticities and $\bar{n}$ is the galaxy surface density, whose values are $0.09$ and $30 \: \mathrm{arcmin}^{-2}$ \citep{Euclid_2020} respectively. The theoretical model used to fit the data is the Limber approximation already presented in Eqs. \eqref{eq:limber}, \eqref{eq:kaiser}, with $\Sigma_\mathrm{mg}(\chi)=1$ since we are now dealing with regular GR lensing.
The optimization can be performed on all 6 \LCDM{} parameters at once, or on a subset of those parameters, while keeping the rest fixed to the values of the fiducial cosmology used to produce the ray-traced spectrum. The ray-traced power spectra that we fit to, are obtained with the fiducial D3A cosmology of FLAMINGO (see Table \ref{tab:cosmology_parameters} for the explicit values of the parameters) in the L2p8 box, in particular the mean of the 8 virtual observers present in the box, in order to minimize the effect of cosmic variance. The MG DMO spectrum is obtained using the overdensity screening method\footnote{The halo screening method cannot be used in DMO simulations, since the halo positions in the lightcones are only available in the hydrodynamical ones.} with $\delta_\mathrm{thresh}=5$ and $\Sigma_0=0.044$.
We show our results in Figure \ref{fig:best_fit_parameters} for the matter density parameter today, $\Omega_\mathrm{m}$, and for $\sigma_8$, as these are the two most relevant parameters when studying the  weak lensing convergence power spectrum. The plot shows a best fit obtained with a cut on the power spectra at $\ell_\mathrm{max}=1500$, corresponding to the pessimistic scale cut for \textit{Euclid} \citep{Euclid_2020}. We see that, in all cases, $\Omega_\mathrm{m}$ is recovered well in accordance with the fiducial value. However, this is not the case for $\sigma_8$, which is approximately $2\sigma$ away for the hydrodynamical spectrum, and reaching more than $5\sigma$ for the MG spectrum. Moreover, the shift in $\sigma_8$ is an overestimation with respect to the fiducial value, as further confirmation that MG models with positive $\Sigma_0$ are not compatible with cosmologies that have a low value for $\sigma_8$. 

This result is an indication that MG effects on the lensing prescription can be more important than baryonic effects when not considered in the analysis of the weak lensing convergence power spectrum in the context of MG extensions to \LCDM{}. This is also expected to depend on the scale cut in the analysis.

\section{Conclusions}
\label{sec:conclusions}

We studied the relevance of modified gravity effects on the weak lensing convergence power spectrum. We used the outputs of the FLAMINGO hydrodynamical simulations as the input for a backward ray-tracing algorithm, corrected to describe a non trivial modification to the lensing equation through the phenomenological $\Sigma_\mathrm{mg}$ function (Eq. \ref{eq:lensing}). We implemented an analytical time dependence for $\Sigma_\mathrm{mg}$ (Eq. \ref{eq:Sigma}) and we implemented scale dependence phenomenologically through two different screening methods, one based on halo positions and one based on over-dense regions. We found that, with both methods, the discrepancy with respect to the standard GR lensing prescription reaches around 5\% at the largest scales (for our chosen value of $\Sigma_0=0.044$, motivated by the latest DESI constraints) decreasing towards smaller scales, where screening becomes important. The two methods lead to similar behavior on large scales, while they differ on small scales, where the halo method shows a dip and regain of power (Figure \ref{fig:cls_delta_vs_haloes}). This is due to the halo method imprinting a characteristic scale at which screening is most efficient (Figure \ref{fig:ratios_source_delta}).

We studied how the power spectrum of the weak lensing convergence depends on the parameters of our model, in particular on the screening halo mass (Figure \ref{fig:cls_ratios_screening_haloes}) and density thresholds (Figure \ref{fig:cls_ratios_screening_delta}), and on $\Sigma_0$, which encodes the amplitude of the deviation of the lensing equation from GR today (Figure \ref{fig:cls_sigma0}). We compared these variations to other known variations, such as cosmological models (Figure \ref{fig:cls_cosmo_vs_MG}) and baryonic feedback (Figure \ref{fig:cls_hydro_vs_MG}), finding that MG effects are comparable with these.

Finally, we fit the MG lensing convergence power spectrum obtained from our pipeline assuming a standard GR \LCDM{} model, to obtain a quantitative indication of the bias in the \LCDM{} parameters resulting from ignoring MG (Figure \ref{fig:best_fit_parameters}). We found that, while $\Omega_\mathrm{m}$ is well recovered, the shift in $\sigma_8$ reaches a bias of over $5\sigma$ with respect to the fiducial value. This is much greater than the bias we get by fitting a hydrodynamical spectrum with a DMO model (for our fiducial scale cut of $\ell_\mathrm{max}=1500$). This result indicates that, when dealing with MG extensions to the \LCDM{} model, MG effects can be more important than baryonic effects.

This work is a first indication that modified lensing should become a standard procedure when analyzing cosmological simulations which include MG. We plan to continue investigating these effects and to study the full case with non-trivial MG modifications to both clustering and lensing in future work.

\section*{Acknowledgements}

MP, MS and AS acknowledge support from the NWO and the Dutch Ministry of Education, Culture and Science (OCW) (through ENW-XL Grant OCENW.XL21.XL21.025 DUSC). AS acknowledges  support from the European Research Council under the H2020 ERC Consolidator Grant “Gravitational Physics from the Universe Large scales Evolution” (Grant No. 101126217 — GraviPULSE). This work used the DiRAC@Durham facility managed by the Institute for Computational Cosmology on behalf of
the STFC DiRAC HPC Facility (www.dirac.ac.uk). The equipment was funded by BEIS capital funding via STFC capital grants
ST/K00042X/1, ST/P002293/1, ST/R002371/1 and ST/S002502/1,
Durham University and STFC operations grant ST/R000832/1.
DiRAC is part of the National e-Infrastructure.

\section*{Data Availability}

The FLAMINGO simulation data will be made publicly available. The base version of the ray-tracing code is available at \url{https://github.com/JegerBroxterman/Lensing_raytrace_FLAMINGO}. The data used to produce the results of this paper is available on reasonable request to the corresponding author.



\bibliographystyle{mnras}
\bibliography{biblio} 

@ARTICLE{Broxterman_2024,
       author = {{Broxterman}, Jeger C. and {Schaller}, Matthieu and {Schaye}, Joop and {Hoekstra}, Henk and {Kuijken}, Konrad and {Helly}, John C. and {Kugel}, Roi and {Braspenning}, Joey and {Elbers}, Willem and {Frenk}, Carlos S. and {Kwan}, Juliana and {McCarthy}, Ian G. and {Salcido}, Jaime and {van Daalen}, Marcel P. and {Vandenbroucke}, Bert},
        title = "{The FLAMINGO project: baryonic impact on weak gravitational lensing convergence peak counts}",
      journal = {\mnras},
         year = 2024,
        month = apr,
       volume = {529},
       number = {3},
        pages = {2309-2326},
          doi = {10.1093/mnras/stae698},
archivePrefix = {arXiv},
       eprint = {2312.08450},
 primaryClass = {astro-ph.CO},
       adsurl = {https://ui.adsabs.harvard.edu/abs/2024MNRAS.529.2309B}
}

@ARTICLE{Silvestri:2009hh,
       author = {{Silvestri}, Alessandra and {Trodden}, Mark},
        title = "{Approaches to understanding cosmic acceleration}",
      journal = {Reports on Progress in Physics},
         year = 2009,
        month = sep,
       volume = {72},
       number = {9},
          eid = {096901},
        pages = {096901},
          doi = {10.1088/0034-4885/72/9/096901},
archivePrefix = {arXiv},
       eprint = {0904.0024},
 primaryClass = {astro-ph.CO},
       adsurl = {https://ui.adsabs.harvard.edu/abs/2009RPPh...72i6901S}
}

@ARTICLE{Kilbinger_2015,
       author = {{Kilbinger}, Martin},
        title = "{Cosmology with cosmic shear observations: a review}",
      journal = {Reports on Progress in Physics},
         year = 2015,
        month = jul,
       volume = {78},
       number = {8},
          eid = {086901},
        pages = {086901},
          doi = {10.1088/0034-4885/78/8/086901},
archivePrefix = {arXiv},
       eprint = {1411.0115},
 primaryClass = {astro-ph.CO},
       adsurl = {https://ui.adsabs.harvard.edu/abs/2015RPPh...78h6901K}
}

@ARTICLE{Hilbert_2009,
       author = {{Hilbert}, S. and {Hartlap}, J. and {White}, S.~D.~M. and {Schneider}, P.},
        title = "{Ray-tracing through the Millennium Simulation: Born corrections and lens-lens coupling in cosmic shear and galaxy-galaxy lensing}",
      journal = {\aap},
         year = 2009,
        month = may,
       volume = {499},
       number = {1},
        pages = {31-43},
          doi = {10.1051/0004-6361/200811054},
archivePrefix = {arXiv},
       eprint = {0809.5035},
 primaryClass = {astro-ph},
       adsurl = {https://ui.adsabs.harvard.edu/abs/2009A&A...499...31H}
}

@ARTICLE{Gorski_2005,
       author = {{G{\'o}rski}, K.~M. and {Hivon}, E. and {Banday}, A.~J. and {Wandelt}, B.~D. and {Hansen}, F.~K. and {Reinecke}, M. and {Bartelmann}, M.},
        title = "{HEALPix: A Framework for High-Resolution Discretization and Fast Analysis of Data Distributed on the Sphere}",
      journal = {\apj},
         year = 2005,
        month = apr,
       volume = {622},
       number = {2},
        pages = {759-771},
          doi = {10.1086/427976},
archivePrefix = {arXiv},
       eprint = {astro-ph/0409513},
 primaryClass = {astro-ph},
       adsurl = {https://ui.adsabs.harvard.edu/abs/2005ApJ...622..759G}
}

@ARTICLE{Schneider_2016,
       author = {{Schneider}, Peter},
        title = "{Generalized shear-ratio tests: A new relation between cosmological distances, and a diagnostic for a redshift-dependent multiplicative bias in shear measurements}",
      journal = {\aap},
         year = 2016,
        month = aug,
       volume = {592},
          eid = {L6},
        pages = {L6},
          doi = {10.1051/0004-6361/201628506},
archivePrefix = {arXiv},
       eprint = {1603.04226},
 primaryClass = {astro-ph.CO},
       adsurl = {https://ui.adsabs.harvard.edu/abs/2016A&A...592L...6S}
}

@ARTICLE{Euclid_2020,
       author = {{Euclid Collaboration} and {Blanchard}, A. and {Camera}, S. and {Carbone}, C. and {Cardone}, V.~F. and {Casas}, S. and {Clesse}, S. and {Ili{\'c}}, S. and {Kilbinger}, M. and {Kitching}, T. and {Kunz}, M. and {Lacasa}, F. and {Linder}, E. and {Majerotto}, E. and {Markovi{\v{c}}}, K. and {Martinelli}, M. and {Pettorino}, V. and {Pourtsidou}, A. and {Sakr}, Z. and {S{\'a}nchez}, A.~G. and {Sapone}, D. and {Tutusaus}, I. and {Yahia-Cherif}, S. and {Yankelevich}, V. and {Andreon}, S. and {Aussel}, H. and {Balaguera-Antol{\'\i}nez}, A. and {Baldi}, M. and {Bardelli}, S. and {Bender}, R. and {Biviano}, A. and {Bonino}, D. and {Boucaud}, A. and {Bozzo}, E. and {Branchini}, E. and {Brau-Nogue}, S. and {Brescia}, M. and {Brinchmann}, J. and {Burigana}, C. and {Cabanac}, R. and {Capobianco}, V. and {Cappi}, A. and {Carretero}, J. and {Carvalho}, C.~S. and {Casas}, R. and {Castander}, F.~J. and {Castellano}, M. and {Cavuoti}, S. and {Cimatti}, A. and {Cledassou}, R. and {Colodro-Conde}, C. and {Congedo}, G. and {Conselice}, C.~J. and {Conversi}, L. and {Copin}, Y. and {Corcione}, L. and {Coupon}, J. and {Courtois}, H.~M. and {Cropper}, M. and {Da Silva}, A. and {de la Torre}, S. and {Di Ferdinando}, D. and {Dubath}, F. and {Ducret}, F. and {Duncan}, C.~A.~J. and {Dupac}, X. and {Dusini}, S. and {Fabbian}, G. and {Fabricius}, M. and {Farrens}, S. and {Fosalba}, P. and {Fotopoulou}, S. and {Fourmanoit}, N. and {Frailis}, M. and {Franceschi}, E. and {Franzetti}, P. and {Fumana}, M. and {Galeotta}, S. and {Gillard}, W. and {Gillis}, B. and {Giocoli}, C. and {G{\'o}mez-Alvarez}, P. and {Graci{\'a}-Carpio}, J. and {Grupp}, F. and {Guzzo}, L. and {Hoekstra}, H. and {Hormuth}, F. and {Israel}, H. and {Jahnke}, K. and {Keihanen}, E. and {Kermiche}, S. and {Kirkpatrick}, C.~C. and {Kohley}, R. and {Kubik}, B. and {Kurki-Suonio}, H. and {Ligori}, S. and {Lilje}, P.~B. and {Lloro}, I. and {Maino}, D. and {Maiorano}, E. and {Marggraf}, O. and {Martinet}, N. and {Marulli}, F. and {Massey}, R. and {Medinaceli}, E. and {Mei}, S. and {Mellier}, Y. and {Metcalf}, B. and {Metge}, J.~J. and {Meylan}, G. and {Moresco}, M. and {Moscardini}, L. and {Munari}, E. and {Nichol}, R.~C. and {Niemi}, S. and {Nucita}, A.~A. and {Padilla}, C. and {Paltani}, S. and {Pasian}, F. and {Percival}, W.~J. and {Pires}, S. and {Polenta}, G. and {Poncet}, M. and {Pozzetti}, L. and {Racca}, G.~D. and {Raison}, F. and {Renzi}, A. and {Rhodes}, J. and {Romelli}, E. and {Roncarelli}, M. and {Rossetti}, E. and {Saglia}, R. and {Schneider}, P. and {Scottez}, V. and {Secroun}, A. and {Sirri}, G. and {Stanco}, L. and {Starck}, J.-L. and {Sureau}, F. and {Tallada-Cresp{\'\i}}, P. and {Tavagnacco}, D. and {Taylor}, A.~N. and {Tenti}, M. and {Tereno}, I. and {Toledo-Moreo}, R. and {Torradeflot}, F. and {Valenziano}, L. and {Vassallo}, T. and {Verdoes Kleijn}, G.~A. and {Viel}, M. and {Wang}, Y. and {Zacchei}, A. and {Zoubian}, J. and {Zucca}, E.},
        title = "{Euclid preparation. VII. Forecast validation for Euclid cosmological probes}",
      journal = {\aap},
         year = 2020,
        month = oct,
       volume = {642},
          eid = {A191},
        pages = {A191},
          doi = {10.1051/0004-6361/202038071},
archivePrefix = {arXiv},
       eprint = {1910.09273},
 primaryClass = {astro-ph.CO},
       adsurl = {https://ui.adsabs.harvard.edu/abs/2020A&A...642A.191E}
}

@ARTICLE{becker2012calclensweaklensingsimulations,
       author = {{Becker}, Matthew R.},
        title = "{CALCLENS: weak lensing simulations for large-area sky surveys and second-order effects in cosmic shear power spectra}",
      journal = {\mnras},
         year = 2013,
        month = oct,
       volume = {435},
       number = {1},
        pages = {115-132},
          doi = {10.1093/mnras/stt1352},
       adsurl = {https://ui.adsabs.harvard.edu/abs/2013MNRAS.435..115B}
}

@ARTICLE{Schaye_2023,
       author = {{Schaye}, Joop and {Kugel}, Roi and {Schaller}, Matthieu and {Helly}, John C. and {Braspenning}, Joey and {Elbers}, Willem and {McCarthy}, Ian G. and {van Daalen}, Marcel P. and {Vandenbroucke}, Bert and {Frenk}, Carlos S. and {Kwan}, Juliana and {Salcido}, Jaime and {Bah{\'e}}, Yannick M. and {Borrow}, Josh and {Chaikin}, Evgenii and {Hahn}, Oliver and {Hu{\v{s}}ko}, Filip and {Jenkins}, Adrian and {Lacey}, Cedric G. and {Nobels}, Folkert S.~J.},
        title = "{The FLAMINGO project: cosmological hydrodynamical simulations for large-scale structure and galaxy cluster surveys}",
      journal = {\mnras},
         year = 2023,
        month = dec,
       volume = {526},
       number = {4},
        pages = {4978-5020},
          doi = {10.1093/mnras/stad2419},
archivePrefix = {arXiv},
       eprint = {2306.04024},
 primaryClass = {astro-ph.CO},
       adsurl = {https://ui.adsabs.harvard.edu/abs/2023MNRAS.526.4978S}
}

@ARTICLE{kugel2023flamingocalibratinglargecosmological,
       author = {{Kugel}, Roi and {Schaye}, Joop and {Schaller}, Matthieu and {Helly}, John C. and {Braspenning}, Joey and {Elbers}, Willem and {Frenk}, Carlos S. and {McCarthy}, Ian G. and {Kwan}, Juliana and {Salcido}, Jaime and {van Daalen}, Marcel P. and {Vandenbroucke}, Bert and {Bah{\'e}}, Yannick M. and {Borrow}, Josh and {Chaikin}, Evgenii and {Hu{\v{s}}ko}, Filip and {Jenkins}, Adrian and {Lacey}, Cedric G. and {Nobels}, Folkert S.~J. and {Vernon}, Ian},
        title = "{FLAMINGO: calibrating large cosmological hydrodynamical simulations with machine learning}",
      journal = {\mnras},
         year = 2023,
        month = dec,
       volume = {526},
       number = {4},
        pages = {6103-6127},
          doi = {10.1093/mnras/stad2540},
archivePrefix = {arXiv},
       eprint = {2306.05492},
 primaryClass = {astro-ph.CO},
       adsurl = {https://ui.adsabs.harvard.edu/abs/2023MNRAS.526.6103K}
}

@ARTICLE{hoyland2025fastgenerationweaklensing,
       author = {{Hoyland}, Sophie and {Winther}, Hans A. and {Saadeh}, Daniela and {Koyama}, Kazuya and {Izard}, Albert},
        title = "{Fast generation of weak lensing maps in modified gravity with COLA}",
      journal = {\mnras},
         year = 2025,
        month = aug,
       volume = {541},
       number = {4},
        pages = {3167-3183},
          doi = {10.1093/mnras/staf1071},
archivePrefix = {arXiv},
       eprint = {2502.14851},
 primaryClass = {astro-ph.CO},
       adsurl = {https://ui.adsabs.harvard.edu/abs/2025MNRAS.541.3167H}
}

@ARTICLE{Schaller_2024,
       author = {{Schaller}, Matthieu and {Borrow}, Josh and {Draper}, Peter W. and {Ivkovic}, Mladen and {McAlpine}, Stuart and {Vandenbroucke}, Bert and {Bah{\'e}}, Yannick and {Chaikin}, Evgenii and {Chalk}, Aidan B.~G. and {Chan}, Tsang Keung and {Correa}, Camila and {van Daalen}, Marcel and {Elbers}, Willem and {Gonnet}, Pedro and {Hausammann}, Lo{\"\i}c and {Helly}, John and {Hu{\v{s}}ko}, Filip and {Kegerreis}, Jacob A. and {Nobels}, Folkert S.~J. and {Ploeckinger}, Sylvia and {Revaz}, Yves and {Roper}, William J. and {Ruiz-Bonilla}, Sergio and {Sandnes}, Thomas D. and {Uyttenhove}, Yolan and {Willis}, James S. and {Xiang}, Zhen},
        title = "{SWIFT: A modern highly-parallel gravity and smoothed particle hydrodynamics solver for astrophysical and cosmological applications}",
      journal = {\mnras},
         year = 2024,
        month = may,
       volume = {530},
       number = {2},
        pages = {2378-2419},
          doi = {10.1093/mnras/stae922},
archivePrefix = {arXiv},
       eprint = {2305.13380},
 primaryClass = {astro-ph.IM},
       adsurl = {https://ui.adsabs.harvard.edu/abs/2024MNRAS.530.2378S}
}

@ARTICLE{Price_2012,
       author = {{Price}, Daniel J.},
        title = "{Smoothed particle hydrodynamics and magnetohydrodynamics}",
      journal = {Journal of Computational Physics},
         year = 2012,
        month = feb,
       volume = {231},
       number = {3},
        pages = {759-794},
          doi = {10.1016/j.jcp.2010.12.011},
archivePrefix = {arXiv},
       eprint = {1012.1885},
 primaryClass = {astro-ph.IM},
       adsurl = {https://ui.adsabs.harvard.edu/abs/2012JCoPh.231..759P}
}

@ARTICLE{Elbers_2021,
       author = {{Elbers}, Willem and {Frenk}, Carlos S. and {Jenkins}, Adrian and {Li}, Baojiu and {Pascoli}, Silvia},
        title = "{An optimal non-linear method for simulating relic neutrinos}",
      journal = {\mnras},
         year = 2021,
        month = oct,
       volume = {507},
       number = {2},
        pages = {2614-2631},
          doi = {10.1093/mnras/stab2260},
archivePrefix = {arXiv},
       eprint = {2010.07321},
 primaryClass = {astro-ph.CO},
       adsurl = {https://ui.adsabs.harvard.edu/abs/2021MNRAS.507.2614E}
}

@ARTICLE{Borrow_2021,
       author = {{Borrow}, Josh and {Schaller}, Matthieu and {Bower}, Richard G. and {Schaye}, Joop},
        title = "{SPHENIX: smoothed particle hydrodynamics for the next generation of galaxy formation simulations}",
      journal = {\mnras},
         year = 2022,
        month = apr,
       volume = {511},
       number = {2},
        pages = {2367-2389},
          doi = {10.1093/mnras/stab3166},
archivePrefix = {arXiv},
       eprint = {2012.03974},
 primaryClass = {astro-ph.GA},
       adsurl = {https://ui.adsabs.harvard.edu/abs/2022MNRAS.511.2367B}
}

@ARTICLE{Booth_2009,
       author = {{Booth}, C.~M. and {Schaye}, Joop},
        title = "{Cosmological simulations of the growth of supermassive black holes and feedback from active galactic nuclei: method and tests}",
      journal = {\mnras},
         year = 2009,
        month = sep,
       volume = {398},
       number = {1},
        pages = {53-74},
          doi = {10.1111/j.1365-2966.2009.15043.x},
archivePrefix = {arXiv},
       eprint = {0904.2572},
 primaryClass = {astro-ph.CO},
       adsurl = {https://ui.adsabs.harvard.edu/abs/2009MNRAS.398...53B}
}

@ARTICLE{Ploeckinger_2020,
       author = {{Ploeckinger}, Sylvia and {Schaye}, Joop},
        title = "{Radiative cooling rates, ion fractions, molecule abundances, and line emissivities including self-shielding and both local and metagalactic radiation fields}",
      journal = {\mnras},
         year = 2020,
        month = oct,
       volume = {497},
       number = {4},
        pages = {4857-4883},
          doi = {10.1093/mnras/staa2172},
archivePrefix = {arXiv},
       eprint = {2006.14322},
 primaryClass = {astro-ph.GA},
       adsurl = {https://ui.adsabs.harvard.edu/abs/2020MNRAS.497.4857P}
}

@ARTICLE{Wiersma_2009,
       author = {{Wiersma}, Robert P.~C. and {Schaye}, Joop and {Theuns}, Tom and {Dalla Vecchia}, Claudio and {Tornatore}, Luca},
        title = "{Chemical enrichment in cosmological, smoothed particle hydrodynamics simulations}",
      journal = {\mnras},
         year = 2009,
        month = oct,
       volume = {399},
       number = {2},
        pages = {574-600},
          doi = {10.1111/j.1365-2966.2009.15331.x},
archivePrefix = {arXiv},
       eprint = {0902.1535},
 primaryClass = {astro-ph.CO},
       adsurl = {https://ui.adsabs.harvard.edu/abs/2009MNRAS.399..574W}
}

@ARTICLE{Dalla_Vecchia_2008,
       author = {{Dalla Vecchia}, Claudio and {Schaye}, Joop},
        title = "{Simulating galactic outflows with kinetic supernova feedback}",
      journal = {\mnras},
         year = 2008,
        month = jul,
       volume = {387},
       number = {4},
        pages = {1431-1444},
          doi = {10.1111/j.1365-2966.2008.13322.x},
archivePrefix = {arXiv},
       eprint = {0801.2770},
 primaryClass = {astro-ph},
       adsurl = {https://ui.adsabs.harvard.edu/abs/2008MNRAS.387.1431D}
}

@ARTICLE{Chaikin_2023,
       author = {{Chaikin}, Evgenii and {Schaye}, Joop and {Schaller}, Matthieu and {Ben{\'\i}tez-Llambay}, Alejandro and {Nobels}, Folkert S.~J. and {Ploeckinger}, Sylvia},
        title = "{A thermal-kinetic subgrid model for supernova feedback in simulations of galaxy formation}",
      journal = {\mnras},
         year = 2023,
        month = aug,
       volume = {523},
       number = {3},
        pages = {3709-3731},
          doi = {10.1093/mnras/stad1626},
archivePrefix = {arXiv},
       eprint = {2211.04619},
 primaryClass = {astro-ph.GA},
       adsurl = {https://ui.adsabs.harvard.edu/abs/2023MNRAS.523.3709C}
}

@ARTICLE{Di_Matteo_2005,
       author = {{Di Matteo}, Tiziana and {Springel}, Volker and {Hernquist}, Lars},
        title = "{Energy input from quasars regulates the growth and activity of black holes and their host galaxies}",
      journal = {\nat},
         year = 2005,
        month = feb,
       volume = {433},
       number = {7026},
        pages = {604-607},
          doi = {10.1038/nature03335},
archivePrefix = {arXiv},
       eprint = {astro-ph/0502199},
 primaryClass = {astro-ph},
       adsurl = {https://ui.adsabs.harvard.edu/abs/2005Natur.433..604D}
}

@ARTICLE{Bahe_2022,
       author = {{Bah{\'e}}, Yannick M. and {Schaye}, Joop and {Schaller}, Matthieu and {Bower}, Richard G. and {Borrow}, Josh and {Chaikin}, Evgenii and {Kugel}, Roi and {Nobels}, Folkert and {Ploeckinger}, Sylvia},
        title = "{The importance of black hole repositioning for galaxy formation simulations}",
      journal = {\mnras},
         year = 2022,
        month = oct,
       volume = {516},
       number = {1},
        pages = {167-184},
          doi = {10.1093/mnras/stac1339},
archivePrefix = {arXiv},
       eprint = {2109.01489},
 primaryClass = {astro-ph.GA},
       adsurl = {https://ui.adsabs.harvard.edu/abs/2022MNRAS.516..167B}
}

@ARTICLE{Abbott_2022,
       author = {{Abbott}, T.~M.~C. and {Aguena}, M. and {Alarcon}, A. and {Allam}, S. and {Alves}, O. and {Amon}, A. and {Andrade-Oliveira}, F. and {Annis}, J. and {Avila}, S. and {Bacon}, D. and {Baxter}, E. and {Bechtol}, K. and {Becker}, M.~R. and {Bernstein}, G.~M. and {Bhargava}, S. and {Birrer}, S. and {Blazek}, J. and {Brandao-Souza}, A. and {Bridle}, S.~L. and {Brooks}, D. and {Buckley-Geer}, E. and {Burke}, D.~L. and {Camacho}, H. and {Campos}, A. and {Carnero Rosell}, A. and {Carrasco Kind}, M. and {Carretero}, J. and {Castander}, F.~J. and {Cawthon}, R. and {Chang}, C. and {Chen}, A. and {Chen}, R. and {Choi}, A. and {Conselice}, C. and {Cordero}, J. and {Costanzi}, M. and {Crocce}, M. and {da Costa}, L.~N. and {da Silva Pereira}, M.~E. and {Davis}, C. and {Davis}, T.~M. and {De Vicente}, J. and {DeRose}, J. and {Desai}, S. and {Di Valentino}, E. and {Diehl}, H.~T. and {Dietrich}, J.~P. and {Dodelson}, S. and {Doel}, P. and {Doux}, C. and {Drlica-Wagner}, A. and {Eckert}, K. and {Eifler}, T.~F. and {Elsner}, F. and {Elvin-Poole}, J. and {Everett}, S. and {Evrard}, A.~E. and {Fang}, X. and {Farahi}, A. and {Fernandez}, E. and {Ferrero}, I. and {Fert{\'e}}, A. and {Fosalba}, P. and {Friedrich}, O. and {Frieman}, J. and {Garc{\'\i}a-Bellido}, J. and {Gatti}, M. and {Gaztanaga}, E. and {Gerdes}, D.~W. and {Giannantonio}, T. and {Giannini}, G. and {Gruen}, D. and {Gruendl}, R.~A. and {Gschwend}, J. and {Gutierrez}, G. and {Harrison}, I. and {Hartley}, W.~G. and {Herner}, K. and {Hinton}, S.~R. and {Hollowood}, D.~L. and {Honscheid}, K. and {Hoyle}, B. and {Huff}, E.~M. and {Huterer}, D. and {Jain}, B. and {James}, D.~J. and {Jarvis}, M. and {Jeffrey}, N. and {Jeltema}, T. and {Kovacs}, A. and {Krause}, E. and {Kron}, R. and {Kuehn}, K. and {Kuropatkin}, N. and {Lahav}, O. and {Leget}, P.-F. and {Lemos}, P. and {Liddle}, A.~R. and {Lidman}, C. and {Lima}, M. and {Lin}, H. and {MacCrann}, N. and {Maia}, M.~A.~G. and {Marshall}, J.~L. and {Martini}, P. and {McCullough}, J. and {Melchior}, P. and {Mena-Fern{\'a}ndez}, J. and {Menanteau}, F. and {Miquel}, R. and {Mohr}, J.~J. and {Morgan}, R. and {Muir}, J. and {Myles}, J. and {Nadathur}, S. and {Navarro-Alsina}, A. and {Nichol}, R.~C. and {Ogando}, R.~L.~C. and {Omori}, Y. and {Palmese}, A. and {Pandey}, S. and {Park}, Y. and {Paz-Chinch{\'o}n}, F. and {Petravick}, D. and {Pieres}, A. and {Plazas Malag{\'o}n}, A.~A. and {Porredon}, A. and {Prat}, J. and {Raveri}, M. and {Rodriguez-Monroy}, M. and {Rollins}, R.~P. and {Romer}, A.~K. and {Roodman}, A. and {Rosenfeld}, R. and {Ross}, A.~J. and {Rykoff}, E.~S. and {Samuroff}, S. and {S{\'a}nchez}, C. and {Sanchez}, E. and {Sanchez}, J. and {Sanchez Cid}, D. and {Scarpine}, V. and {Schubnell}, M. and {Scolnic}, D. and {Secco}, L.~F. and {Serrano}, S. and {Sevilla-Noarbe}, I. and {Sheldon}, E. and {Shin}, T. and {Smith}, M. and {Soares-Santos}, M. and {Suchyta}, E. and {Swanson}, M.~E.~C. and {Tabbutt}, M. and {Tarle}, G. and {Thomas}, D. and {To}, C. and {Troja}, A. and {Troxel}, M.~A. and {Tucker}, D.~L. and {Tutusaus}, I. and {Varga}, T.~N. and {Walker}, A.~R. and {Weaverdyck}, N. and {Wechsler}, R. and {Weller}, J. and {Yanny}, B. and {Yin}, B. and {Zhang}, Y. and {Zuntz}, J. and {DES Collaboration}},
        title = "{Dark Energy Survey Year 3 results: Cosmological constraints from galaxy clustering and weak lensing}",
      journal = {\prd},
         year = 2022,
        month = jan,
       volume = {105},
       number = {2},
          eid = {023520},
        pages = {023520},
          doi = {10.1103/PhysRevD.105.023520},
archivePrefix = {arXiv},
       eprint = {2105.13549},
 primaryClass = {astro-ph.CO},
       adsurl = {https://ui.adsabs.harvard.edu/abs/2022PhRvD.105b3520A}
}

@ARTICLE{Amon_2022,
       author = {{Amon}, A. and {Robertson}, N.~C. and {Miyatake}, H. and {Heymans}, C. and {White}, M. and {DeRose}, J. and {Yuan}, S. and {Wechsler}, R.~H. and {Varga}, T.~N. and {Bocquet}, S. and {Dvornik}, A. and {More}, S. and {Ross}, A.~J. and {Hoekstra}, H. and {Alarcon}, A. and {Asgari}, M. and {Blazek}, J. and {Campos}, A. and {Chen}, R. and {Choi}, A. and {Crocce}, M. and {Diehl}, H.~T. and {Doux}, C. and {Eckert}, K. and {Elvin-Poole}, J. and {Everett}, S. and {Fert{\'e}}, A. and {Gatti}, M. and {Giannini}, G. and {Gruen}, D. and {Gruendl}, R.~A. and {Hartley}, W.~G. and {Herner}, K. and {Hildebrandt}, H. and {Huang}, S. and {Huff}, E.~M. and {Joachimi}, B. and {Lee}, S. and {MacCrann}, N. and {Myles}, J. and {Navarro-Alsina}, A. and {Nishimichi}, T. and {Prat}, J. and {Secco}, L.~F. and {Sevilla-Noarbe}, I. and {Sheldon}, E. and {Shin}, T. and {Tr{\"o}ster}, T. and {Troxel}, M.~A. and {Tutusaus}, I. and {Wright}, A.~H. and {Yin}, B. and {Aguena}, M. and {Allam}, S. and {Annis}, J. and {Bacon}, D. and {Bilicki}, M. and {Brooks}, D. and {Burke}, D.~L. and {Carnero Rosell}, A. and {Carretero}, J. and {Castander}, F.~J. and {Cawthon}, R. and {Costanzi}, M. and {da Costa}, L.~N. and {Pereira}, M.~E.~S. and {de Jong}, J. and {De Vicente}, J. and {Desai}, S. and {Dietrich}, J.~P. and {Doel}, P. and {Ferrero}, I. and {Frieman}, J. and {Garc{\'\i}a-Bellido}, J. and {Gerdes}, D.~W. and {Gschwend}, J. and {Gutierrez}, G. and {Hinton}, S.~R. and {Hollowood}, D.~L. and {Honscheid}, K. and {Huterer}, D. and {Kannawadi}, A. and {Kuehn}, K. and {Kuropatkin}, N. and {Lahav}, O. and {Lima}, M. and {Maia}, M.~A.~G. and {Marshall}, J.~L. and {Menanteau}, F. and {Miquel}, R. and {Mohr}, J.~J. and {Morgan}, R. and {Muir}, J. and {Paz-Chinch{\'o}n}, F. and {Pieres}, A. and {Plazas Malag{\'o}n}, A.~A. and {Porredon}, A. and {Rodriguez-Monroy}, M. and {Roodman}, A. and {Sanchez}, E. and {Serrano}, S. and {Shan}, H. and {Suchyta}, E. and {Swanson}, M.~E.~C. and {Tarle}, G. and {Thomas}, D. and {To}, C. and {Zhang}, Y.},
        title = "{Consistent lensing and clustering in a low-S$_{8}$ Universe with BOSS, DES Year 3, HSC Year 1, and KiDS-1000}",
      journal = {\mnras},
         year = 2023,
        month = jan,
       volume = {518},
       number = {1},
        pages = {477-503},
          doi = {10.1093/mnras/stac2938},
archivePrefix = {arXiv},
       eprint = {2202.07440},
 primaryClass = {astro-ph.CO},
       adsurl = {https://ui.adsabs.harvard.edu/abs/2023MNRAS.518..477A}
}

@ARTICLE{Planck_2020,
       author = {{Planck Collaboration} and {Aghanim}, N. and {Akrami}, Y. and {Ashdown}, M. and {Aumont}, J. and {Baccigalupi}, C. and {Ballardini}, M. and {Banday}, A.~J. and {Barreiro}, R.~B. and {Bartolo}, N. and {Basak}, S. and {Battye}, R. and {Benabed}, K. and {Bernard}, J.-P. and {Bersanelli}, M. and {Bielewicz}, P. and {Bock}, J.~J. and {Bond}, J.~R. and {Borrill}, J. and {Bouchet}, F.~R. and {Boulanger}, F. and {Bucher}, M. and {Burigana}, C. and {Butler}, R.~C. and {Calabrese}, E. and {Cardoso}, J.-F. and {Carron}, J. and {Challinor}, A. and {Chiang}, H.~C. and {Chluba}, J. and {Colombo}, L.~P.~L. and {Combet}, C. and {Contreras}, D. and {Crill}, B.~P. and {Cuttaia}, F. and {de Bernardis}, P. and {de Zotti}, G. and {Delabrouille}, J. and {Delouis}, J.-M. and {Di Valentino}, E. and {Diego}, J.~M. and {Dor{\'e}}, O. and {Douspis}, M. and {Ducout}, A. and {Dupac}, X. and {Dusini}, S. and {Efstathiou}, G. and {Elsner}, F. and {En{\ss}lin}, T.~A. and {Eriksen}, H.~K. and {Fantaye}, Y. and {Farhang}, M. and {Fergusson}, J. and {Fernandez-Cobos}, R. and {Finelli}, F. and {Forastieri}, F. and {Frailis}, M. and {Fraisse}, A.~A. and {Franceschi}, E. and {Frolov}, A. and {Galeotta}, S. and {Galli}, S. and {Ganga}, K. and {G{\'e}nova-Santos}, R.~T. and {Gerbino}, M. and {Ghosh}, T. and {Gonz{\'a}lez-Nuevo}, J. and {G{\'o}rski}, K.~M. and {Gratton}, S. and {Gruppuso}, A. and {Gudmundsson}, J.~E. and {Hamann}, J. and {Handley}, W. and {Hansen}, F.~K. and {Herranz}, D. and {Hildebrandt}, S.~R. and {Hivon}, E. and {Huang}, Z. and {Jaffe}, A.~H. and {Jones}, W.~C. and {Karakci}, A. and {Keih{\"a}nen}, E. and {Keskitalo}, R. and {Kiiveri}, K. and {Kim}, J. and {Kisner}, T.~S. and {Knox}, L. and {Krachmalnicoff}, N. and {Kunz}, M. and {Kurki-Suonio}, H. and {Lagache}, G. and {Lamarre}, J.-M. and {Lasenby}, A. and {Lattanzi}, M. and {Lawrence}, C.~R. and {Le Jeune}, M. and {Lemos}, P. and {Lesgourgues}, J. and {Levrier}, F. and {Lewis}, A. and {Liguori}, M. and {Lilje}, P.~B. and {Lilley}, M. and {Lindholm}, V. and {L{\'o}pez-Caniego}, M. and {Lubin}, P.~M. and {Ma}, Y.-Z. and {Mac{\'\i}as-P{\'e}rez}, J.~F. and {Maggio}, G. and {Maino}, D. and {Mandolesi}, N. and {Mangilli}, A. and {Marcos-Caballero}, A. and {Maris}, M. and {Martin}, P.~G. and {Martinelli}, M. and {Mart{\'\i}nez-Gonz{\'a}lez}, E. and {Matarrese}, S. and {Mauri}, N. and {McEwen}, J.~D. and {Meinhold}, P.~R. and {Melchiorri}, A. and {Mennella}, A. and {Migliaccio}, M. and {Millea}, M. and {Mitra}, S. and {Miville-Desch{\^e}nes}, M.-A. and {Molinari}, D. and {Montier}, L. and {Morgante}, G. and {Moss}, A. and {Natoli}, P. and {N{\o}rgaard-Nielsen}, H.~U. and {Pagano}, L. and {Paoletti}, D. and {Partridge}, B. and {Patanchon}, G. and {Peiris}, H.~V. and {Perrotta}, F. and {Pettorino}, V. and {Piacentini}, F. and {Polastri}, L. and {Polenta}, G. and {Puget}, J.-L. and {Rachen}, J.~P. and {Reinecke}, M. and {Remazeilles}, M. and {Renzi}, A. and {Rocha}, G. and {Rosset}, C. and {Roudier}, G. and {Rubi{\~n}o-Mart{\'\i}n}, J.~A. and {Ruiz-Granados}, B. and {Salvati}, L. and {Sandri}, M. and {Savelainen}, M. and {Scott}, D. and {Shellard}, E.~P.~S. and {Sirignano}, C. and {Sirri}, G. and {Spencer}, L.~D. and {Sunyaev}, R. and {Suur-Uski}, A.-S. and {Tauber}, J.~A. and {Tavagnacco}, D. and {Tenti}, M. and {Toffolatti}, L. and {Tomasi}, M. and {Trombetti}, T. and {Valenziano}, L. and {Valiviita}, J. and {Van Tent}, B. and {Vibert}, L. and {Vielva}, P. and {Villa}, F. and {Vittorio}, N. and {Wandelt}, B.~D. and {Wehus}, I.~K. and {White}, M. and {White}, S.~D.~M. and {Zacchei}, A. and {Zonca}, A.},
        title = "{Planck 2018 results. VI. Cosmological parameters}",
      journal = {\aap},
         year = 2020,
        month = sep,
       volume = {641},
          eid = {A6},
        pages = {A6},
          doi = {10.1051/0004-6361/201833910},
archivePrefix = {arXiv},
       eprint = {1807.06209},
 primaryClass = {astro-ph.CO},
       adsurl = {https://ui.adsabs.harvard.edu/abs/2020A&A...641A...6P}
}

@ARTICLE{McGibbon_2025,
       author = {{McGibbon}, Robert and {Helly}, John and {Schaye}, Joop and {Schaller}, Matthieu and {Vandenbroucke}, Bert},
        title = "{SOAP: A Python Package for Calculating the Properties of Galaxies and Halos Formed in Cosmological Simulations}",
      journal = {The Journal of Open Source Software},
         year = 2025,
        month = jul,
       volume = {10},
       number = {111},
          eid = {8252},
        pages = {8252},
          doi = {10.21105/joss.08252},
archivePrefix = {arXiv},
       eprint = {2507.22669},
 primaryClass = {astro-ph.IM},
       adsurl = {https://ui.adsabs.harvard.edu/abs/2025JOSS...10.8252M}
}

@ARTICLE{albuquerque2025euclid,
       author = {{Euclid Collaboration} and {Albuquerque}, I.~S. and {Frusciante}, N. and {Sakr}, Z. and {Srinivasan}, S. and {Atayde}, L. and {Bose}, B. and {Cardone}, V.~F. and {Casas}, S. and {Martinelli}, M. and {Noller}, J. and {Teixeira}, E.~M. and {Thomas}, D.~B. and {Tutusaus}, I. and {Cataneo}, M. and {Koyama}, K. and {Lombriser}, L. and {Pace}, F. and {Silvestri}, A. and {Aghanim}, N. and {Amara}, A. and {Andreon}, S. and {Auricchio}, N. and {Baccigalupi}, C. and {Baldi}, M. and {Bardelli}, S. and {Biviano}, A. and {Bonino}, D. and {Branchini}, E. and {Brescia}, M. and {Brinchmann}, J. and {Camera}, S. and {Ca{\~n}as-Herrera}, G. and {Capobianco}, V. and {Carbone}, C. and {Carretero}, J. and {Castellano}, M. and {Castignani}, G. and {Cavuoti}, S. and {Chambers}, K.~C. and {Cimatti}, A. and {Colodro-Conde}, C. and {Congedo}, G. and {Conselice}, C.~J. and {Conversi}, L. and {Copin}, Y. and {Corcione}, L. and {Courbin}, F. and {Courtois}, H.~M. and {Da Silva}, A. and {Degaudenzi}, H. and {de la Torre}, S. and {De Lucia}, G. and {Di Giorgio}, A.~M. and {Dole}, H. and {Dubath}, F. and {Duncan}, C.~A.~J. and {Dupac}, X. and {Dusini}, S. and {Ealet}, A. and {Escoffier}, S. and {Farina}, M. and {Farrens}, S. and {Faustini}, F. and {Ferriol}, S. and {Finelli}, F. and {Fosalba}, P. and {Fotopoulou}, S. and {Frailis}, M. and {Franceschi}, E. and {Fumana}, M. and {Galeotta}, S. and {Gillis}, B. and {Giocoli}, C. and {Gracia-Carpio}, J. and {Grazian}, A. and {Grupp}, F. and {Guzzo}, L. and {Haugan}, S.~V.~H. and {Holmes}, W. and {Hormuth}, F. and {Hornstrup}, A. and {Hudelot}, P. and {Ili{\'c}}, S. and {Jahnke}, K. and {Jhabvala}, M. and {Joachimi}, B. and {Keih{\"a}nen}, E. and {Kermiche}, S. and {Kiessling}, A. and {Kilbinger}, M. and {Kubik}, B. and {Kunz}, M. and {Kurki-Suonio}, H. and {Le Brun}, A.~M.~C. and {Ligori}, S. and {Lilje}, P.~B. and {Lindholm}, V. and {Lloro}, I. and {Mainetti}, G. and {Maino}, D. and {Maiorano}, E. and {Mansutti}, O. and {Marggraf}, O. and {Markovic}, K. and {Martinet}, N. and {Marulli}, F. and {Massey}, R. and {Medinaceli}, E. and {Mei}, S. and {Mellier}, Y. and {Meneghetti}, M. and {Merlin}, E. and {Meylan}, G. and {Mora}, A. and {Moresco}, M. and {Moscardini}, L. and {Nichol}, R.~C. and {Niemi}, S.-M. and {Nightingale}, J.~W. and {Padilla}, C. and {Paltani}, S. and {Pasian}, F. and {Percival}, W.~J. and {Pettorino}, V. and {Pires}, S. and {Polenta}, G. and {Poncet}, M. and {Popa}, L.~A. and {Pozzetti}, L. and {Raison}, F. and {Rebolo}, R. and {Renzi}, A. and {Rhodes}, J. and {Riccio}, G. and {Romelli}, E. and {Roncarelli}, M. and {Saglia}, R. and {Sapone}, D. and {Sartoris}, B. and {Schewtschenko}, J.~A. and {Schirmer}, M. and {Schrabback}, T. and {Secroun}, A. and {Sefusatti}, E. and {Seidel}, G. and {Serrano}, S. and {Sirignano}, C. and {Sirri}, G. and {Stanco}, L. and {Steinwagner}, J. and {Tallada-Cresp{\'\i}}, P. and {Taylor}, A.~N. and {Tereno}, I. and {Tessore}, N. and {Toft}, S. and {Toledo-Moreo}, R. and {Torradeflot}, F. and {Valentijn}, E.~A. and {Valenziano}, L. and {Valiviita}, J. and {Vassallo}, T. and {Verdoes Kleijn}, G. and {Veropalumbo}, A. and {Wang}, Y. and {Weller}, J. and {Zamorani}, G. and {Zucca}, E. and {Bozzo}, E. and {Burigana}, C. and {Calabrese}, M. and {Di Ferdinando}, D. and {Escartin Vigo}, J.~A. and {Matthew}, S. and {Mauri}, N. and {Pezzotta}, A. and {P{\"o}ntinen}, M. and {Porciani}, C. and {Scottez}, V. and {Tenti}, M. and {Viel}, M. and {Wiesmann}, M. and {Akrami}, Y. and {Allevato}, V. and {Alvi}, S. and {Anselmi}, S. and {Archidiacono}, M. and {Atrio-Barandela}, F. and {Balaguera-Antolinez}, A. and {Ballardini}, M. and {Bertacca}, D. and {Blanchard}, A. and {Blot}, L. and {Borgani}, S. and {Brown}, M.~L. and {Bruton}, S. and {Cabanac}, R. and {Calabro}, A. and {Camacho Quevedo}, B. and {Cappi}, A.},
        title = "{Euclid preparation. Constraining parameterised models of modifications of gravity with the spectroscopic and photometric primary probes}",
      journal = {arXiv e-prints},
         year = 2025,
        month = jun,
          eid = {arXiv:2506.03008},
        pages = {arXiv:2506.03008},
          doi = {10.48550/arXiv.2506.03008},
archivePrefix = {arXiv},
       eprint = {2506.03008},
 primaryClass = {astro-ph.CO},
       adsurl = {https://ui.adsabs.harvard.edu/abs/2025arXiv250603008E}
}

@ARTICLE{lsstsciencecollaboration2009lsstsciencebookversion,
       author = {{LSST Science Collaboration} and {Abell}, Paul A. and {Allison}, Julius and {Anderson}, Scott F. and {Andrew}, John R. and {Angel}, J. Roger P. and {Armus}, Lee and {Arnett}, David and {Asztalos}, S.~J. and {Axelrod}, Tim S. and {Bailey}, Stephen and {Ballantyne}, D.~R. and {Bankert}, Justin R. and {Barkhouse}, Wayne A. and {Barr}, Jeffrey D. and {Barrientos}, L. Felipe and {Barth}, Aaron J. and {Bartlett}, James G. and {Becker}, Andrew C. and {Becla}, Jacek and {Beers}, Timothy C. and {Bernstein}, Joseph P. and {Biswas}, Rahul and {Blanton}, Michael R. and {Bloom}, Joshua S. and {Bochanski}, John J. and {Boeshaar}, Pat and {Borne}, Kirk D. and {Bradac}, Marusa and {Brandt}, W.~N. and {Bridge}, Carrie R. and {Brown}, Michael E. and {Brunner}, Robert J. and {Bullock}, James S. and {Burgasser}, Adam J. and {Burge}, James H. and {Burke}, David L. and {Cargile}, Phillip A. and {Chandrasekharan}, Srinivasan and {Chartas}, George and {Chesley}, Steven R. and {Chu}, You-Hua and {Cinabro}, David and {Claire}, Mark W. and {Claver}, Charles F. and {Clowe}, Douglas and {Connolly}, A.~J. and {Cook}, Kem H. and {Cooke}, Jeff and {Cooray}, Asantha and {Covey}, Kevin R. and {Culliton}, Christopher S. and {de Jong}, Roelof and {de Vries}, Willem H. and {Debattista}, Victor P. and {Delgado}, Francisco and {Dell'Antonio}, Ian P. and {Dhital}, Saurav and {Di Stefano}, Rosanne and {Dickinson}, Mark and {Dilday}, Benjamin and {Djorgovski}, S.~G. and {Dobler}, Gregory and {Donalek}, Ciro and {Dubois-Felsmann}, Gregory and {Durech}, Josef and {Eliasdottir}, Ardis and {Eracleous}, Michael and {Eyer}, Laurent and {Falco}, Emilio E. and {Fan}, Xiaohui and {Fassnacht}, Christopher D. and {Ferguson}, Harry C. and {Fernandez}, Yanga R. and {Fields}, Brian D. and {Finkbeiner}, Douglas and {Figueroa}, Eduardo E. and {Fox}, Derek B. and {Francke}, Harold and {Frank}, James S. and {Frieman}, Josh and {Fromenteau}, Sebastien and {Furqan}, Muhammad and {Galaz}, Gaspar and {Gal-Yam}, A. and {Garnavich}, Peter and {Gawiser}, Eric and {Geary}, John and {Gee}, Perry and {Gibson}, Robert R. and {Gilmore}, Kirk and {Grace}, Emily A. and {Green}, Richard F. and {Gressler}, William J. and {Grillmair}, Carl J. and {Habib}, Salman and {Haggerty}, J.~S. and {Hamuy}, Mario and {Harris}, Alan W. and {Hawley}, Suzanne L. and {Heavens}, Alan F. and {Hebb}, Leslie and {Henry}, Todd J. and {Hileman}, Edward and {Hilton}, Eric J. and {Hoadley}, Keri and {Holberg}, J.~B. and {Holman}, Matt J. and {Howell}, Steve B. and {Infante}, Leopoldo and {Ivezic}, Zeljko and {Jacoby}, Suzanne H. and {Jain}, Bhuvnesh and {R} and {Jedicke} and {Jee}, M. James and {Garrett Jernigan}, J. and {Jha}, Saurabh W. and {Johnston}, Kathryn V. and {Jones}, R. Lynne and {Juric}, Mario and {Kaasalainen}, Mikko and {Styliani} and {Kafka} and {Kahn}, Steven M. and {Kaib}, Nathan A. and {Kalirai}, Jason and {Kantor}, Jeff and {Kasliwal}, Mansi M. and {Keeton}, Charles R. and {Kessler}, Richard and {Knezevic}, Zoran and {Kowalski}, Adam and {Krabbendam}, Victor L. and {Krughoff}, K. Simon and {Kulkarni}, Shrinivas and {Kuhlman}, Stephen and {Lacy}, Mark and {Lepine}, Sebastien and {Liang}, Ming and {Lien}, Amy and {Lira}, Paulina and {Long}, Knox S. and {Lorenz}, Suzanne and {Lotz}, Jennifer M. and {Lupton}, R.~H. and {Lutz}, Julie and {Macri}, Lucas M. and {Mahabal}, Ashish A. and {Mandelbaum}, Rachel and {Marshall}, Phil and {May}, Morgan and {McGehee}, Peregrine M. and {Meadows}, Brian T. and {Meert}, Alan and {Milani}, Andrea and {Miller}, Christopher J. and {Miller}, Michelle and {Mills}, David and {Minniti}, Dante and {Monet}, David and {Mukadam}, Anjum S. and {Nakar}, Ehud and {Neill}, Douglas R. and {Newman}, Jeffrey A. and {Nikolaev}, Sergei and {Nordby}, Martin and {O'Connor}, Paul and {Oguri}, Masamune and {Oliver}, John and {Olivier}, Scot S. and {Olsen}, Julia K. and {Olsen}, Knut and {Olszewski}, Edward W. and {Oluseyi}, Hakeem and {Padilla}, Nelson D. and {Parker}, Alex and {Pepper}, Joshua and {Peterson}, John R. and {Petry}, Catherine and {Pinto}, Philip A. and {Pizagno}, James L. and {Popescu}, Bogdan and {Prsa}, Andrej and {Radcka}, Veljko and {Raddick}, M. Jordan and {Rasmussen}, Andrew and {Rau}, Arne and {Rho}, Jeonghee and {Rhoads}, James E. and {Richards}, Gordon T. and {Ridgway}, Stephen T. and {Robertson}, Brant E. and {Roskar}, Rok and {Saha}, Abhijit and {Sarajedini}, Ata and {Scannapieco}, Evan and {Schalk}, Terry and {Schindler}, Rafe and {Schmidt}, Samuel},
        title = "{LSST Science Book, Version 2.0}",
      journal = {arXiv e-prints},
         year = 2009,
        month = dec,
          eid = {arXiv:0912.0201},
        pages = {arXiv:0912.0201},
          doi = {10.48550/arXiv.0912.0201},
archivePrefix = {arXiv},
       eprint = {0912.0201},
 primaryClass = {astro-ph.IM},
       adsurl = {https://ui.adsabs.harvard.edu/abs/2009arXiv0912.0201L}
}

@ARTICLE{spergel2015widefieldinfrarredsurveytelescopeastrophysics,
       author = {{Spergel}, D. and {Gehrels}, N. and {Baltay}, C. and {Bennett}, D. and {Breckinridge}, J. and {Donahue}, M. and {Dressler}, A. and {Gaudi}, B.~S. and {Greene}, T. and {Guyon}, O. and {Hirata}, C. and {Kalirai}, J. and {Kasdin}, N.~J. and {Macintosh}, B. and {Moos}, W. and {Perlmutter}, S. and {Postman}, M. and {Rauscher}, B. and {Rhodes}, J. and {Wang}, Y. and {Weinberg}, D. and {Benford}, D. and {Hudson}, M. and {Jeong}, W.-S. and {Mellier}, Y. and {Traub}, W. and {Yamada}, T. and {Capak}, P. and {Colbert}, J. and {Masters}, D. and {Penny}, M. and {Savransky}, D. and {Stern}, D. and {Zimmerman}, N. and {Barry}, R. and {Bartusek}, L. and {Carpenter}, K. and {Cheng}, E. and {Content}, D. and {Dekens}, F. and {Demers}, R. and {Grady}, K. and {Jackson}, C. and {Kuan}, G. and {Kruk}, J. and {Melton}, M. and {Nemati}, B. and {Parvin}, B. and {Poberezhskiy}, I. and {Peddie}, C. and {Ruffa}, J. and {Wallace}, J.~K. and {Whipple}, A. and {Wollack}, E. and {Zhao}, F.},
        title = "{Wide-Field InfrarRed Survey Telescope-Astrophysics Focused Telescope Assets WFIRST-AFTA 2015 Report}",
      journal = {arXiv e-prints},
         year = 2015,
        month = mar,
          eid = {arXiv:1503.03757},
        pages = {arXiv:1503.03757},
          doi = {10.48550/arXiv.1503.03757},
archivePrefix = {arXiv},
       eprint = {1503.03757},
 primaryClass = {astro-ph.IM},
       adsurl = {https://ui.adsabs.harvard.edu/abs/2015arXiv150303757S}
}

@ARTICLE{Khoury_2004,
       author = {{Khoury}, Justin and {Weltman}, Amanda},
        title = "{Chameleon cosmology}",
      journal = {\prd},
         year = 2004,
        month = feb,
       volume = {69},
       number = {4},
          eid = {044026},
        pages = {044026},
          doi = {10.1103/PhysRevD.69.044026},
archivePrefix = {arXiv},
       eprint = {astro-ph/0309411},
 primaryClass = {astro-ph},
       adsurl = {https://ui.adsabs.harvard.edu/abs/2004PhRvD..69d4026K}
}

@ARTICLE{Hinterbichler_2010,
       author = {{Hinterbichler}, Kurt and {Khoury}, Justin},
        title = "{Screening Long-Range Forces through Local Symmetry Restoration}",
      journal = {\prl},
         year = 2010,
        month = jun,
       volume = {104},
       number = {23},
          eid = {231301},
        pages = {231301},
          doi = {10.1103/PhysRevLett.104.231301},
archivePrefix = {arXiv},
       eprint = {1001.4525},
 primaryClass = {hep-th},
       adsurl = {https://ui.adsabs.harvard.edu/abs/2010PhRvL.104w1301H}
}

@ARTICLE{Nicolis_2009,
       author = {{Nicolis}, Alberto and {Rattazzi}, Riccardo and {Trincherini}, Enrico},
        title = "{Galileon as a local modification of gravity}",
      journal = {\prd},
         year = 2009,
        month = mar,
       volume = {79},
       number = {6},
          eid = {064036},
        pages = {064036},
          doi = {10.1103/PhysRevD.79.064036},
archivePrefix = {arXiv},
       eprint = {0811.2197},
 primaryClass = {hep-th},
       adsurl = {https://ui.adsabs.harvard.edu/abs/2009PhRvD..79f4036N}
}

@ARTICLE{BABICHEV_2009,
       author = {{Babichev}, E. and {Deffayet}, C. and {Ziour}, R.},
        title = "{k-MOUFLAGE Gravity}",
      journal = {International Journal of Modern Physics D},
         year = 2009,
        month = jan,
       volume = {18},
       number = {14},
        pages = {2147-2154},
          doi = {10.1142/S0218271809016107},
archivePrefix = {arXiv},
       eprint = {0905.2943},
 primaryClass = {hep-th},
       adsurl = {https://ui.adsabs.harvard.edu/abs/2009IJMPD..18.2147B}
}

@ARTICLE{Joyce_2015,
       author = {{Joyce}, Austin and {Jain}, Bhuvnesh and {Khoury}, Justin and {Trodden}, Mark},
        title = "{Beyond the cosmological standard model}",
      journal = {\physrep},
         year = 2015,
        month = mar,
       volume = {568},
        pages = {1-98},
          doi = {10.1016/j.physrep.2014.12.002},
archivePrefix = {arXiv},
       eprint = {1407.0059},
 primaryClass = {astro-ph.CO},
       adsurl = {https://ui.adsabs.harvard.edu/abs/2015PhR...568....1J}
}

@ARTICLE{Anderson_2014,
       author = {{Anderson}, Lauren and {Aubourg}, {\'E}ric and {Bailey}, Stephen and {Beutler}, Florian and {Bhardwaj}, Vaishali and {Blanton}, Michael and {Bolton}, Adam S. and {Brinkmann}, J. and {Brownstein}, Joel R. and {Burden}, Angela and {Chuang}, Chia-Hsun and {Cuesta}, Antonio J. and {Dawson}, Kyle S. and {Eisenstein}, Daniel J. and {Escoffier}, Stephanie and {Gunn}, James E. and {Guo}, Hong and {Ho}, Shirley and {Honscheid}, Klaus and {Howlett}, Cullan and {Kirkby}, David and {Lupton}, Robert H. and {Manera}, Marc and {Maraston}, Claudia and {McBride}, Cameron K. and {Mena}, Olga and {Montesano}, Francesco and {Nichol}, Robert C. and {Nuza}, Sebasti{\'a}n E. and {Olmstead}, Matthew D. and {Padmanabhan}, Nikhil and {Palanque-Delabrouille}, Nathalie and {Parejko}, John and {Percival}, Will J. and {Petitjean}, Patrick and {Prada}, Francisco and {Price-Whelan}, Adrian M. and {Reid}, Beth and {Roe}, Natalie A. and {Ross}, Ashley J. and {Ross}, Nicholas P. and {Sabiu}, Cristiano G. and {Saito}, Shun and {Samushia}, Lado and {S{\'a}nchez}, Ariel G. and {Schlegel}, David J. and {Schneider}, Donald P. and {Scoccola}, Claudia G. and {Seo}, Hee-Jong and {Skibba}, Ramin A. and {Strauss}, Michael A. and {Swanson}, Molly E.~C. and {Thomas}, Daniel and {Tinker}, Jeremy L. and {Tojeiro}, Rita and {Maga{\~n}a}, Mariana Vargas and {Verde}, Licia and {Wake}, David A. and {Weaver}, Benjamin A. and {Weinberg}, David H. and {White}, Martin and {Xu}, Xiaoying and {Y{\`e}che}, Christophe and {Zehavi}, Idit and {Zhao}, Gong-Bo},
        title = "{The clustering of galaxies in the SDSS-III Baryon Oscillation Spectroscopic Survey: baryon acoustic oscillations in the Data Releases 10 and 11 Galaxy samples}",
      journal = {\mnras},
         year = 2014,
        month = jun,
       volume = {441},
       number = {1},
        pages = {24-62},
          doi = {10.1093/mnras/stu523},
archivePrefix = {arXiv},
       eprint = {1312.4877},
 primaryClass = {astro-ph.CO},
       adsurl = {https://ui.adsabs.harvard.edu/abs/2014MNRAS.441...24A}
}

@ARTICLE{Will_2014,
       author = {{Will}, Clifford M.},
        title = "{The Confrontation between General Relativity and Experiment}",
      journal = {Living Reviews in Relativity},
         year = 2014,
        month = dec,
       volume = {17},
       number = {1},
          eid = {4},
        pages = {4},
          doi = {10.12942/lrr-2014-4},
archivePrefix = {arXiv},
       eprint = {1403.7377},
 primaryClass = {gr-qc},
       adsurl = {https://ui.adsabs.harvard.edu/abs/2014LRR....17....4W}
}

@ARTICLE{Semboloni_2011,
       author = {{Semboloni}, Elisabetta and {Hoekstra}, Henk and {Schaye}, Joop and {van Daalen}, Marcel P. and {McCarthy}, Ian G.},
        title = "{Quantifying the effect of baryon physics on weak lensing tomography}",
      journal = {\mnras},
         year = 2011,
        month = nov,
       volume = {417},
       number = {3},
        pages = {2020-2035},
          doi = {10.1111/j.1365-2966.2011.19385.x},
archivePrefix = {arXiv},
       eprint = {1105.1075},
 primaryClass = {astro-ph.CO},
       adsurl = {https://ui.adsabs.harvard.edu/abs/2011MNRAS.417.2020S}
}

@ARTICLE{Verde_2019,
       author = {{Verde}, Licia and {Treu}, Tommaso and {Riess}, Adam G.},
        title = "{Tensions between the early and late Universe}",
      journal = {Nature Astronomy},
         year = 2019,
        month = sep,
       volume = {3},
        pages = {891-895},
          doi = {10.1038/s41550-019-0902-0},
archivePrefix = {arXiv},
       eprint = {1907.10625},
 primaryClass = {astro-ph.CO},
       adsurl = {https://ui.adsabs.harvard.edu/abs/2019NatAs...3..891V}
}

@ARTICLE{Di_Valentino_2021,
       author = {{Di Valentino}, Eleonora and {Anchordoqui}, Luis A. and {Akarsu}, {\"O}zg{\"u}r and {Ali-Haimoud}, Yacine and {Amendola}, Luca and {Arendse}, Nikki and {Asgari}, Marika and {Ballardini}, Mario and {Basilakos}, Spyros and {Battistelli}, Elia and {Benetti}, Micol and {Birrer}, Simon and {Bouchet}, Fran{\c{c}}ois R. and {Bruni}, Marco and {Calabrese}, Erminia and {Camarena}, David and {Capozziello}, Salvatore and {Chen}, Angela and {Chluba}, Jens and {Chudaykin}, Anton and {Colg{\'a}in}, Eoin {\'O}. and {Cyr-Racine}, Francis-Yan and {de Bernardis}, Paolo and {de Cruz P{\'e}rez}, Javier and {Delabrouille}, Jacques and {Dunkley}, Jo and {Escamilla-Rivera}, Celia and {Fert{\'e}}, Agn{\`e}s and {Finelli}, Fabio and {Freedman}, Wendy and {Frusciante}, Noemi and {Giusarma}, Elena and {G{\'o}mez-Valent}, Adri{\`a} and {Handley}, Will and {Harrison}, Ian and {Hart}, Luke and {Heavens}, Alan and {Hildebrandt}, Hendrik and {Holz}, Daniel and {Huterer}, Dragan and {Ivanov}, Mikhail M. and {Joudaki}, Shahab and {Kamionkowski}, Marc and {Karwal}, Tanvi and {Knox}, Lloyd and {Kumar}, Suresh and {Lamagna}, Luca and {Lesgourgues}, Julien and {Lucca}, Matteo and {Marra}, Valerio and {Masi}, Silvia and {Matarrese}, Sabino and {Mazumdar}, Arindam and {Melchiorri}, Alessandro and {Mena}, Olga and {Mersini-Houghton}, Laura and {Miranda}, Vivian and {Moreno-Pulido}, Cristian and {Mota}, David F. and {Muir}, Jessica and {Mukherjee}, Ankan and {Niedermann}, Florian and {Notari}, Alessio and {Nunes}, Rafael C. and {Pace}, Francesco and {Paliathanasis}, Andronikos and {Palmese}, Antonella and {Pan}, Supriya and {Paoletti}, Daniela and {Pettorino}, Valeria and {Piacentini}, Francesco and {Poulin}, Vivian and {Raveri}, Marco and {Riess}, Adam G. and {Salzano}, Vincenzo and {Saridakis}, Emmanuel N. and {Sen}, Anjan A. and {Shafieloo}, Arman and {Shajib}, Anowar J. and {Silk}, Joseph and {Silvestri}, Alessandra and {Sloth}, Martin S. and {Smith}, Tristan L. and {Sol{\`a} Peracaula}, Joan and {van de Bruck}, Carsten and {Verde}, Licia and {Visinelli}, Luca and {Wandelt}, Benjamin D. and {Wang}, Deng and {Wang}, Jian-Min and {Yadav}, Anil K. and {Yang}, Weiqiang},
        title = "{Cosmology Intertwined III: f{\ensuremath{\sigma}}$_{8}$ and S$_{8}$}",
      journal = {Astroparticle Physics},
         year = 2021,
        month = sep,
       volume = {131},
          eid = {102604},
        pages = {102604},
          doi = {10.1016/j.astropartphys.2021.102604},
archivePrefix = {arXiv},
       eprint = {2008.11285},
 primaryClass = {astro-ph.CO},
       adsurl = {https://ui.adsabs.harvard.edu/abs/2021APh...13102604D}
}

@ARTICLE{Bartelmann_2001,
       author = {{Bartelmann}, M. and {Schneider}, P.},
        title = "{Weak gravitational lensing}",
      journal = {\physrep},
         year = 2001,
        month = jan,
       volume = {340},
       number = {4-5},
        pages = {291-472},
          doi = {10.1016/S0370-1573(00)00082-X},
archivePrefix = {arXiv},
       eprint = {astro-ph/9912508},
 primaryClass = {astro-ph},
       adsurl = {https://ui.adsabs.harvard.edu/abs/2001PhR...340..291B}
}

@ARTICLE{Koyama_2016,
       author = {{Koyama}, Kazuya},
        title = "{Cosmological tests of modified gravity}",
      journal = {Reports on Progress in Physics},
         year = 2016,
        month = apr,
       volume = {79},
       number = {4},
          eid = {046902},
        pages = {046902},
          doi = {10.1088/0034-4885/79/4/046902},
archivePrefix = {arXiv},
       eprint = {1504.04623},
 primaryClass = {astro-ph.CO},
       adsurl = {https://ui.adsabs.harvard.edu/abs/2016RPPh...79d6902K}
}

@ARTICLE{Chisari_2018,
       author = {{Chisari}, N.~E. and {Richardson}, M.~L.~A. and {Devriendt}, J. and {Dubois}, Y. and {Schneider}, A. and {Le Brun}, A.~M.~C. and {Beckmann}, R.~S. and {Peirani}, S. and {Slyz}, A. and {Pichon}, C.},
        title = "{The impact of baryons on the matter power spectrum from the Horizon-AGN cosmological hydrodynamical simulation}",
      journal = {\mnras},
         year = 2018,
        month = nov,
       volume = {480},
       number = {3},
        pages = {3962-3977},
          doi = {10.1093/mnras/sty2093},
archivePrefix = {arXiv},
       eprint = {1801.08559},
 primaryClass = {astro-ph.CO},
       adsurl = {https://ui.adsabs.harvard.edu/abs/2018MNRAS.480.3962C}
}

@ARTICLE{harnoisderaps_2022,
       author = {{Harnois-D{\'e}raps}, Joachim and {Hernandez-Aguayo}, Cesar and {Cuesta-Lazaro}, Carolina and {Arnold}, Christian and {Li}, Baojiu and {Davies}, Christopher T. and {Cai}, Yan-Chuan},
        title = "{MGLENS: Modified gravity weak lensing simulations for emulation-based cosmological inference}",
      journal = {\mnras},
         year = 2023,
        month = nov,
       volume = {525},
       number = {4},
        pages = {6336-6358},
          doi = {10.1093/mnras/stad2700},
archivePrefix = {arXiv},
       eprint = {2211.05779},
 primaryClass = {astro-ph.CO},
       adsurl = {https://ui.adsabs.harvard.edu/abs/2023MNRAS.525.6336H}
}

@ARTICLE{Zhang_2007,
       author = {{Zhang}, Pengjie and {Liguori}, Michele and {Bean}, Rachel and {Dodelson}, Scott},
        title = "{Probing Gravity at Cosmological Scales by Measurements which Test the Relationship between Gravitational Lensing and Matter Overdensity}",
      journal = {\prl},
         year = 2007,
        month = oct,
       volume = {99},
       number = {14},
          eid = {141302},
        pages = {141302},
          doi = {10.1103/PhysRevLett.99.141302},
archivePrefix = {arXiv},
       eprint = {0704.1932},
 primaryClass = {astro-ph},
       adsurl = {https://ui.adsabs.harvard.edu/abs/2007PhRvL..99n1302Z}
}

@ARTICLE{Amendola_2008,
       author = {{Amendola}, Luca and {Kunz}, Martin and {Sapone}, Domenico},
        title = "{Measuring the dark side (with weak lensing)}",
      journal = {\jcap},
         year = 2008,
        month = apr,
       volume = {2008},
       number = {4},
          eid = {013},
        pages = {013},
          doi = {10.1088/1475-7516/2008/04/013},
archivePrefix = {arXiv},
       eprint = {0704.2421},
 primaryClass = {astro-ph},
       adsurl = {https://ui.adsabs.harvard.edu/abs/2008JCAP...04..013A}
}

@ARTICLE{Daniel_2010,
       author = {{Daniel}, Scott F. and {Linder}, Eric V.},
        title = "{Confronting general relativity with further cosmological data}",
      journal = {\prd},
         year = 2010,
        month = nov,
       volume = {82},
       number = {10},
          eid = {103523},
        pages = {103523},
          doi = {10.1103/PhysRevD.82.103523},
archivePrefix = {arXiv},
       eprint = {1008.0397},
 primaryClass = {astro-ph.CO},
       adsurl = {https://ui.adsabs.harvard.edu/abs/2010PhRvD..82j3523D}
}

@ARTICLE{Simpson_2012,
       author = {{Simpson}, Fergus and {Heymans}, Catherine and {Parkinson}, David and {Blake}, Chris and {Kilbinger}, Martin and {Benjamin}, Jonathan and {Erben}, Thomas and {Hildebrandt}, Hendrik and {Hoekstra}, Henk and {Kitching}, Thomas D. and {Mellier}, Yannick and {Miller}, Lance and {Van Waerbeke}, Ludovic and {Coupon}, Jean and {Fu}, Liping and {Harnois-D{\'e}raps}, Joachim and {Hudson}, Michael J. and {Kuijken}, Koenraad and {Rowe}, Barnaby and {Schrabback}, Tim and {Semboloni}, Elisabetta and {Vafaei}, Sanaz and {Velander}, Malin},
        title = "{CFHTLenS: testing the laws of gravity with tomographic weak lensing and redshift-space distortions}",
      journal = {\mnras},
         year = 2013,
        month = mar,
       volume = {429},
       number = {3},
        pages = {2249-2263},
          doi = {10.1093/mnras/sts493},
archivePrefix = {arXiv},
       eprint = {1212.3339},
 primaryClass = {astro-ph.CO},
       adsurl = {https://ui.adsabs.harvard.edu/abs/2013MNRAS.429.2249S}
}

@ARTICLE{planck_2016,
       author = {{Planck Collaboration} and {Ade}, P.~A.~R. and {Aghanim}, N. and {Arnaud}, M. and {Ashdown}, M. and {Aumont}, J. and {Baccigalupi}, C. and {Banday}, A.~J. and {Barreiro}, R.~B. and {Bartolo}, N. and {Battaner}, E. and {Battye}, R. and {Benabed}, K. and {Beno{\^\i}t}, A. and {Benoit-L{\'e}vy}, A. and {Bernard}, J.-P. and {Bersanelli}, M. and {Bielewicz}, P. and {Bock}, J.~J. and {Bonaldi}, A. and {Bonavera}, L. and {Bond}, J.~R. and {Borrill}, J. and {Bouchet}, F.~R. and {Bucher}, M. and {Burigana}, C. and {Butler}, R.~C. and {Calabrese}, E. and {Cardoso}, J.-F. and {Catalano}, A. and {Challinor}, A. and {Chamballu}, A. and {Chiang}, H.~C. and {Christensen}, P.~R. and {Church}, S. and {Clements}, D.~L. and {Colombi}, S. and {Colombo}, L.~P.~L. and {Combet}, C. and {Couchot}, F. and {Coulais}, A. and {Crill}, B.~P. and {Curto}, A. and {Cuttaia}, F. and {Danese}, L. and {Davies}, R.~D. and {Davis}, R.~J. and {de Bernardis}, P. and {de Rosa}, A. and {de Zotti}, G. and {Delabrouille}, J. and {D{\'e}sert}, F.-X. and {Diego}, J.~M. and {Dole}, H. and {Donzelli}, S. and {Dor{\'e}}, O. and {Douspis}, M. and {Ducout}, A. and {Dupac}, X. and {Efstathiou}, G. and {Elsner}, F. and {En{\ss}lin}, T.~A. and {Eriksen}, H.~K. and {Fergusson}, J. and {Finelli}, F. and {Forni}, O. and {Frailis}, M. and {Fraisse}, A.~A. and {Franceschi}, E. and {Frejsel}, A. and {Galeotta}, S. and {Galli}, S. and {Ganga}, K. and {Giard}, M. and {Giraud-H{\'e}raud}, Y. and {Gjerl{\o}w}, E. and {Gonz{\'a}lez-Nuevo}, J. and {G{\'o}rski}, K.~M. and {Gratton}, S. and {Gregorio}, A. and {Gruppuso}, A. and {Gudmundsson}, J.~E. and {Hansen}, F.~K. and {Hanson}, D. and {Harrison}, D.~L. and {Heavens}, A. and {Helou}, G. and {Henrot-Versill{\'e}}, S. and {Hern{\'a}ndez-Monteagudo}, C. and {Herranz}, D. and {Hildebrandt}, S.~R. and {Hivon}, E. and {Hobson}, M. and {Holmes}, W.~A. and {Hornstrup}, A. and {Hovest}, W. and {Huang}, Z. and {Huffenberger}, K.~M. and {Hurier}, G. and {Jaffe}, A.~H. and {Jaffe}, T.~R. and {Jones}, W.~C. and {Juvela}, M. and {Keih{\"a}nen}, E. and {Keskitalo}, R. and {Kisner}, T.~S. and {Knoche}, J. and {Kunz}, M. and {Kurki-Suonio}, H. and {Lagache}, G. and {L{\"a}hteenm{\"a}ki}, A. and {Lamarre}, J.-M. and {Lasenby}, A. and {Lattanzi}, M. and {Lawrence}, C.~R. and {Leonardi}, R. and {Lesgourgues}, J. and {Levrier}, F. and {Lewis}, A. and {Liguori}, M. and {Lilje}, P.~B. and {Linden-V{\o}rnle}, M. and {L{\'o}pez-Caniego}, M. and {Lubin}, P.~M. and {Ma}, Y.-Z. and {Mac{\'\i}as-P{\'e}rez}, J.~F. and {Maggio}, G. and {Maino}, D. and {Mandolesi}, N. and {Mangilli}, A. and {Marchini}, A. and {Maris}, M. and {Martin}, P.~G. and {Martinelli}, M. and {Mart{\'\i}nez-Gonz{\'a}lez}, E. and {Masi}, S. and {Matarrese}, S. and {McGehee}, P. and {Meinhold}, P.~R. and {Melchiorri}, A. and {Mendes}, L. and {Mennella}, A. and {Migliaccio}, M. and {Mitra}, S. and {Miville-Desch{\^e}nes}, M.-A. and {Moneti}, A. and {Montier}, L. and {Morgante}, G. and {Mortlock}, D. and {Moss}, A. and {Munshi}, D. and {Murphy}, J.~A. and {Narimani}, A. and {Naselsky}, P. and {Nati}, F. and {Natoli}, P. and {Netterfield}, C.~B. and {N{\o}rgaard-Nielsen}, H.~U. and {Noviello}, F. and {Novikov}, D. and {Novikov}, I. and {Oxborrow}, C.~A. and {Paci}, F. and {Pagano}, L. and {Pajot}, F. and {Paoletti}, D. and {Pasian}, F. and {Patanchon}, G. and {Pearson}, T.~J. and {Perdereau}, O. and {Perotto}, L. and {Perrotta}, F. and {Pettorino}, V. and {Piacentini}, F. and {Piat}, M. and {Pierpaoli}, E. and {Pietrobon}, D. and {Plaszczynski}, S. and {Pointecouteau}, E. and {Polenta}, G. and {Popa}, L. and {Pratt}, G.~W. and {Pr{\'e}zeau}, G. and {Prunet}, S. and {Puget}, J.-L. and {Rachen}, J.~P. and {Reach}, W.~T. and {Rebolo}, R. and {Reinecke}, M. and {Remazeilles}, M. and {Renault}, C. and {Renzi}, A. and {Ristorcelli}, I. and {Rocha}, G. and {Rosset}, C. and {Rossetti}, M. and {Roudier}, G. and {Rowan-Robinson}, M. and {Rubi{\~n}o-Mart{\'\i}n}, J.~A. and {Rusholme}, B.},
        title = "{Planck 2015 results. XIV. Dark energy and modified gravity}",
      journal = {\aap},
         year = 2016,
        month = sep,
       volume = {594},
          eid = {A14},
        pages = {A14},
          doi = {10.1051/0004-6361/201525814},
archivePrefix = {arXiv},
       eprint = {1502.01590},
 primaryClass = {astro-ph.CO},
       adsurl = {https://ui.adsabs.harvard.edu/abs/2016A&A...594A..14P}
}

@ARTICLE{Lewis_2000,
       author = {{Lewis}, Antony and {Challinor}, Anthony and {Lasenby}, Anthony},
        title = "{Efficient Computation of Cosmic Microwave Background Anisotropies in Closed Friedmann-Robertson-Walker Models}",
      journal = {\apj},
         year = 2000,
        month = aug,
       volume = {538},
       number = {2},
        pages = {473-476},
          doi = {10.1086/309179},
archivePrefix = {arXiv},
       eprint = {astro-ph/9911177},
 primaryClass = {astro-ph},
       adsurl = {https://ui.adsabs.harvard.edu/abs/2000ApJ...538..473L}
}

@ARTICLE{Mead_2021,
       author = {{Mead}, A.~J. and {Brieden}, S. and {Tr{\"o}ster}, T. and {Heymans}, C.},
        title = "{HMCODE-2020: improved modelling of non-linear cosmological power spectra with baryonic feedback}",
      journal = {\mnras},
         year = 2021,
        month = mar,
       volume = {502},
       number = {1},
        pages = {1401-1422},
          doi = {10.1093/mnras/stab082},
archivePrefix = {arXiv},
       eprint = {2009.01858},
 primaryClass = {astro-ph.CO},
       adsurl = {https://ui.adsabs.harvard.edu/abs/2021MNRAS.502.1401M}
}

@ARTICLE{bardeen_1980,
       author = {{Bardeen}, James M.},
        title = "{Gauge-invariant cosmological perturbations}",
      journal = {\prd},
         year = 1980,
        month = oct,
       volume = {22},
       number = {8},
        pages = {1882-1905},
          doi = {10.1103/PhysRevD.22.1882},
       adsurl = {https://ui.adsabs.harvard.edu/abs/1980PhRvD..22.1882B}
}

@ARTICLE{Limber:1954zz,
       author = {{Limber}, D. Nelson},
        title = "{The Analysis of Counts of the Extragalactic Nebulae in Terms of a Fluctuating Density Field. II.}",
      journal = {\apj},
         year = 1954,
        month = may,
       volume = {119},
        pages = {655},
          doi = {10.1086/145870},
       adsurl = {https://ui.adsabs.harvard.edu/abs/1954ApJ...119..655L}
}

@ARTICLE{Loverde_2008,
       author = {{LoVerde}, Marilena and {Afshordi}, Niayesh},
        title = "{Extended Limber approximation}",
      journal = {\prd},
         year = 2008,
        month = dec,
       volume = {78},
       number = {12},
          eid = {123506},
        pages = {123506},
          doi = {10.1103/PhysRevD.78.123506},
archivePrefix = {arXiv},
       eprint = {0809.5112},
 primaryClass = {astro-ph},
       adsurl = {https://ui.adsabs.harvard.edu/abs/2008PhRvD..78l3506L}
}

@ARTICLE{Kaiser:1991qi,
       author = {{Kaiser}, Nick},
        title = "{Weak Gravitational Lensing of Distant Galaxies}",
      journal = {\apj},
         year = 1992,
        month = apr,
       volume = {388},
        pages = {272},
          doi = {10.1086/171151},
       adsurl = {https://ui.adsabs.harvard.edu/abs/1992ApJ...388..272K}
}

@ARTICLE{Clifton_2012,
       author = {{Clifton}, Timothy and {Ferreira}, Pedro G. and {Padilla}, Antonio and {Skordis}, Constantinos},
        title = "{Modified gravity and cosmology}",
      journal = {\physrep},
         year = 2012,
        month = mar,
       volume = {513},
       number = {1},
        pages = {1-189},
          doi = {10.1016/j.physrep.2012.01.001},
archivePrefix = {arXiv},
       eprint = {1106.2476},
 primaryClass = {astro-ph.CO},
       adsurl = {https://ui.adsabs.harvard.edu/abs/2012PhR...513....1C}
}

@ARTICLE{Wang_2023,
       author = {{Wang}, Zhuangfei and {Mirpoorian}, Seyed Hamidreza and {Pogosian}, Levon and {Silvestri}, Alessandra and {Zhao}, Gong-Bo},
        title = "{New MGCAMB tests of gravity with CosmoMC and Cobaya}",
      journal = {\jcap},
         year = 2023,
        month = aug,
       volume = {2023},
       number = {8},
          eid = {038},
        pages = {038},
          doi = {10.1088/1475-7516/2023/08/038},
archivePrefix = {arXiv},
       eprint = {2305.05667},
 primaryClass = {astro-ph.CO},
       adsurl = {https://ui.adsabs.harvard.edu/abs/2023JCAP...08..038W}
}

@ARTICLE{Zhao_2009,
       author = {{Zhao}, Gong-Bo and {Pogosian}, Levon and {Silvestri}, Alessandra and {Zylberberg}, Joel},
        title = "{Searching for modified growth patterns with tomographic surveys}",
      journal = {\prd},
         year = 2009,
        month = apr,
       volume = {79},
       number = {8},
          eid = {083513},
        pages = {083513},
          doi = {10.1103/PhysRevD.79.083513},
archivePrefix = {arXiv},
       eprint = {0809.3791},
 primaryClass = {astro-ph},
       adsurl = {https://ui.adsabs.harvard.edu/abs/2009PhRvD..79h3513Z}
      }

@ARTICLE{Sakr_2022,
       author = {{Sakr}, Z. and {Martinelli}, M.},
        title = "{Cosmological constraints on sub-horizon scales modified gravity theories with MGCLASS II}",
      journal = {\jcap},
         year = 2022,
        month = may,
       volume = {2022},
       number = {5},
          eid = {030},
        pages = {030},
          doi = {10.1088/1475-7516/2022/05/030},
archivePrefix = {arXiv},
       eprint = {2112.14175},
 primaryClass = {astro-ph.CO},
       adsurl = {https://ui.adsabs.harvard.edu/abs/2022JCAP...05..030S}
}

@ARTICLE{DESI_2024VII,
       author = {{DESI Collaboration} and {Adame}, A.~G. and {Aguilar}, J. and {Ahlen}, S. and {Alam}, S. and {Alexander}, D.~M. and {Allende Prieto}, C. and {Alvarez}, M. and {Alves}, O. and {Anand}, A. and {Andrade}, U. and {Armengaud}, E. and {Avila}, S. and {Aviles}, A. and {Awan}, H. and {Bahr-Kalus}, B. and {Bailey}, S. and {Baltay}, C. and {Bault}, A. and {Behera}, J. and {BenZvi}, S. and {Beutler}, F. and {Bianchi}, D. and {Blake}, C. and {Blum}, R. and {Bonici}, M. and {Brieden}, S. and {Brodzeller}, A. and {Brooks}, D. and {Buckley-Geer}, E. and {Burtin}, E. and {Calderon}, R. and {Canning}, R. and {Carnero Rosell}, A. and {Cereskaite}, R. and {Cervantes-Cota}, J.~L. and {Chabanier}, S. and {Chaussidon}, E. and {Chaves-Montero}, J. and {Chebat}, D. and {Chen}, S. and {Chen}, X. and {Claybaugh}, T. and {Cole}, S. and {Cuceu}, A. and {Davis}, T.~M. and {Dawson}, K. and {de la Macorra}, A. and {de Mattia}, A. and {Deiosso}, N. and {Dey}, A. and {Dey}, B. and {Ding}, Z. and {Doel}, P. and {Edelstein}, J. and {Eftekharzadeh}, S. and {Eisenstein}, D.~J. and {Elbers}, W. and {Elliott}, A. and {Fagrelius}, P. and {Fanning}, K. and {Ferraro}, S. and {Ereza}, J. and {Findlay}, N. and {Flaugher}, B. and {Font-Ribera}, A. and {Forero-S{\'a}nchez}, D. and {Forero-Romero}, J.~E. and {Frenk}, C.~S. and {Garcia-Quintero}, C. and {Garrison}, L.~H. and {Gazta{\~n}aga}, E. and {Gil-Mar{\'\i}n}, H. and {Gontcho}, S. Gontcho A. and {Gonzalez-Morales}, A.~X. and {Gonzalez-Perez}, V. and {Gordon}, C. and {Green}, D. and {Gruen}, D. and {Gsponer}, R. and {Gutierrez}, G. and {Guy}, J. and {Hadzhiyska}, B. and {Hahn}, C. and {Hanif}, M.~M.~S. and {Herrera-Alcantar}, H.~K. and {Honscheid}, K. and {Howlett}, C. and {Huterer}, D. and {Ir{\v{s}}i{\v{c}}}, V. and {Ishak}, M. and {Joyce}, R. and {Juneau}, S. and {Kara{\c{c}}ayl{\i}}, N.~G. and {Kehoe}, R. and {Kent}, S. and {Kirkby}, D. and {Kong}, H. and {Koposov}, S.~E. and {Kremin}, A. and {Krolewski}, A. and {Lahav}, O. and {Lai}, Y. and {Lan}, T.-W. and {Landriau}, M. and {Lang}, D. and {Lasker}, J. and {Le Goff}, J.~M. and {Le Guillou}, L. and {Leauthaud}, A. and {Levi}, M.~E. and {Li}, T.~S. and {Lodha}, K. and {Magneville}, C. and {Manera}, M. and {Margala}, D. and {Martini}, P. and {Matthewson}, W. and {Maus}, M. and {McDonald}, P. and {Medina-Varela}, L. and {Meisner}, A. and {Mena-Fern{\'a}ndez}, J. and {Miquel}, R. and {Moon}, J. and {Moore}, S. and {Moustakas}, J. and {Mudur}, N. and {Mueller}, E. and {Mu{\~n}oz-Guti{\'e}rrez}, A. and {Myers}, A.~D. and {Nadathur}, S. and {Napolitano}, L. and {Neveux}, R. and {Newman}, J.~A. and {Nguyen}, N.~M. and {Nie}, J. and {Niz}, G. and {Noriega}, H.~E. and {Padmanabhan}, N. and {Paillas}, E. and {Palanque-Delabrouille}, N. and {Pan}, J. and {Penmetsa}, S. and {Percival}, W.~J. and {Pieri}, M.~M. and {Pinon}, M. and {Poppett}, C. and {Porredon}, A. and {Prada}, F. and {P{\'e}rez-Fern{\'a}ndez}, A. and {P{\'e}rez-R{\`a}fols}, I. and {Rabinowitz}, D. and {Raichoor}, A. and {Ram{\'\i}rez-P{\'e}rez}, C. and {Ramirez-Solano}, S. and {Rashkovetskyi}, M. and {Ravoux}, C. and {Rezaie}, M. and {Rich}, J. and {Rocher}, A. and {Rockosi}, C. and {Roe}, N.~A. and {Rosado-Marin}, A. and {Ross}, A.~J. and {Rossi}, G. and {Ruggeri}, R. and {Ruhlmann-Kleider}, V. and {Samushia}, L. and {Sanchez}, E. and {Saulder}, C. and {Schlafly}, E.~F. and {Schlegel}, D. and {Schubnell}, M. and {Seo}, H. and {Shafieloo}, A. and {Sharples}, R. and {Silber}, J. and {Slosar}, A. and {Smith}, A. and {Sprayberry}, D. and {Tan}, T. and {Tarl{\'e}}, G. and {Taylor}, P. and {Trusov}, S. and {Vaisakh}, R. and {Valcin}, D. and {Valdes}, F. and {Valogiannis}, G. and {Vargas-Maga{\~n}a}, M. and {Verde}, L. and {Walther}, M. and {Wang}, B. and {Wang}, M.~S. and {Weaver}, B.~A. and {Weaverdyck}, N. and {Wechsler}, R.~H. and {Weinberg}, D.~H. and {White}, M. and {Wilson}, M.~J. and {Yi}, L.},
        title = "{DESI 2024 VII: cosmological constraints from the full-shape modeling of clustering measurements}",
      journal = {\jcap},
         year = 2025,
        month = jul,
       volume = {2025},
       number = {7},
          eid = {028},
        pages = {028},
          doi = {10.1088/1475-7516/2025/07/028},
archivePrefix = {arXiv},
       eprint = {2411.12022},
 primaryClass = {astro-ph.CO},
       adsurl = {https://ui.adsabs.harvard.edu/abs/2025JCAP...07..028A}
}

@ARTICLE{Saadeh_2024,
       author = {{Saadeh}, Daniela and {Koyama}, Kazuya and {Morice-Atkinson}, Xan},
        title = "{A field-level emulator for modified gravity}",
      journal = {\mnras},
         year = 2025,
        month = feb,
       volume = {537},
       number = {1},
        pages = {448-463},
          doi = {10.1093/mnras/stae2807},
archivePrefix = {arXiv},
       eprint = {2406.03374},
 primaryClass = {astro-ph.CO},
       adsurl = {https://ui.adsabs.harvard.edu/abs/2025MNRAS.537..448S}
}





\bsp	
\label{lastpage}
\end{document}